\documentclass[twocolumn]{aastex631}

\newcommand{\Bpos}{\ensuremath{\hat{B}_{\rm pos}}}

\newcommand{\sofia}{\emph{SOFIA}}
\newcommand{\hawc}{\emph{HAWC+}}

\newcommand{\hii}{\ion{H}{2}}
\newcommand{\cii}{\ion{C}{2}}
\usepackage{blindtext}

\usepackage{amsmath}

\usepackage{multirow}

\usepackage{needspace}
\usepackage{etoolbox}
\pretocmd{\section}{\needspace{3\baselineskip}}{}{}
\pretocmd{\subsection}{\needspace{3\baselineskip}}{}{}
\pretocmd{\subsubsection}{\needspace{3\baselineskip}}{}{}

\begin{document}

\title{SIMPLIFI -- \underline{S}tudy of \underline{I}nterstellar \underline{M}agnetic \underline{P}olarization: a \underline{L}egacy \underline{I}nvestigation of \underline{Fi}laments. I. \\ Magnetically-Guided Accretion onto the DR21 Ridge}
\shorttitle{SIMPLIFI I: Magnetically-Guided Accretion onto the DR21 Ridge}

\author[0000-0003-2133-4862]{Thushara G.S. Pillai}
\affiliation{Haystack Observatory, Massachusetts Institute of Technology, 99 Millstone Road, Westford, MA 01886, USA}
\email{thushara@mit.edu}

\author[0000-0002-5094-6393]{Jens Kauffmann}
\affiliation{Haystack Observatory, Massachusetts Institute of Technology, 99 Millstone Road, Westford, MA 01886, USA}

\author[0000-0002-0294-4465]{Juan D.\ Soler}
\affiliation{Istituto di Astrofisica e Planetologia Spaziali (IAPS). INAF. Via Fosso del Cavaliere 100, 00133 Roma, Italy}
\affiliation{University of Vienna, Department of Astrophysics, T\"urkenschanzstra{\ss}e 17, 1180 Wien, Austria}

\author[0000-0002-3871-010X]{Mark Heyer}
\affiliation{Department of Astronomy, University of Massachusetts, Amherst, MA 01003, USA}

\author[0000-0002-2885-1806]{Philip C. Myers}
\affiliation{Center for Astrophysics \textbar Harvard and Smithsonian, 60 Garden Street, Cambridge, MA 02138, USA}

\author[0000-0002-4666-609X]{Laura M.\ Fissel}
\affiliation{Department of Physics, Engineering Physics and Astronomy, Queen's University, Kingston, ON K7L 3N6, Canada}

\author[0000-0002-9947-4956]{Dan Clemens}
\affiliation{Institute for Astrophysical Research, Boston University, 725 Commonwealth Avenue, Boston, MA 02215, USA}

\author[0000-0003-3081-6898]{Koji Sugitani}
\affiliation{Graduate School of Science, Nagoya City University, 1~Yamanohata, Mizuho-cho, Mizuho-ku, Nagoya, Aichi 467-8501, Japan}

\author[0000-0001-5357-6538]{Enrique Lopez-Rodriguez}
\affiliation{Department of Physics \& Astronomy, University of South Carolina, Columbia, SC 29208, USA}

\author[0000-0001-5431-2294]{Fumitaka Nakamura}
\affiliation{National Astronomical Observatory of Japan, Mitaka, Tokyo 181-8588, Japan}

\author[0000-0003-3869-6501]{Andrea Giannetti}
\affiliation{INAF - Istituto di Radioastronomia di Bologna, Via Gobetti 101, 40129 Bologna, Italy}

\author[0000-0002-0368-9160]{Daniel Seifried}
\affiliation{Universit\"at zu K\"oln, I.\ Physikalisches Institut, Z\"ulpicher Str.~77, D-50937 K\"oln, Germany}

\author[0000-0002-6622-8396]{Paul F. Goldsmith}
\affiliation{Jet Propulsion Laboratory, California Institute of Technology, 4800 Oak Grove Drive, Pasadena, CA 91109, USA}

\author[0000-0002-5135-8657]{Helmut Wiesemeyer}
\affiliation{Max-Planck-Institut f\"ur Radioastronomie, Auf dem H\"ugel 69, D-53121 Bonn, Germany}

\author[0000-0002-4324-0034]{Evangelia Ntormousi}
\affiliation{Scuola Normale Superiore di Pisa, Italy}

\author[0000-0003-2020-2649]{Gabriel Franco}
\affiliation{Departamento de F\'{\i}sica, ICEx, Universidade Federal de Minas Gerais, CP~702, 30123-970 Belo Horizonte, Brazil}

\author[0000-0001-5222-9139]{Stefan Reissl}
\affiliation{Zentrum f\"ur Astronomie der Universität Heidelberg, Institut für Theoretische Astrophysik, Albert-Ueberle-Str. 2, 69120 Heidelberg, Germany}

\author[0000-0001-6459-0669]{Karl M.\ Menten}
\altaffiliation{Karl~Menten, dearly missed by the SIMPLIFI team, passed away in Dec.~2024.}
\affiliation{Max-Planck-Institut f\"ur Radioastronomie, Auf dem H\"ugel 69, D-53121 Bonn, Germany}

\begin{abstract}

We present first results from SIMPLIFI (\underline{S}tudy of \underline{I}nterstellar \underline{M}agnetic \underline{P}olarization: a \underline{L}egacy \underline{I}nvestigation of \underline{Fi}laments), a \sofia/\hawc\ 214~$\mu$m polarimetric survey of Galactic molecular cloud filaments. We trace magnetic field morphology from the DR21 Main Ridge into surrounding sub-filaments at $\sim{}0.1~\rm{}pc$ resolution, extending polarimetric detections for the first time beyond high-column-density regions probed by prior submillimeter observations. We compare the plane-of-sky orientations of the magnetic field, $\hat{B}_{\rm{}pos}$, the projected gravitational acceleration, $\vec{g}_{\rm{}pos}$, and the intensity gradient rotated by $90\arcdeg$, $\hat{\nabla}_{\perp}I$. The relative orientation of $\hat{B}_{\rm{}pos}$ and $\hat{\nabla}_{\perp}I$ transitions from preferentially parallel in sub-filaments to perpendicular in the DR21 Main Ridge at $N({\rm{}H_2})\sim{}2\times{}10^{22}~\rm{}cm^{-2}$, consistent with thresholds seen with \emph{Planck} and expected in clouds formed from strongly magnetized, sub-Alfv\'enic, magnetically sub-critical gas ($\mathcal{M}_{\rm{}A}\lesssim{}1$, $M/\Phi_B<[M/\Phi_B]_{\rm{}cr}$).

We find that the relative alignment between orientations shows region-to-region and pixel-to-pixel variations at fixed column density. Column density alone is thus not sufficient to encode changes in magnetic field structure. Theoretical models must account for additional drivers.

Our central finding is that $\vec{g}_{\rm{}pos}$ and $\hat{B}_{\rm{}pos}$ remain aligned throughout the cloud regardless of column density or environment, unlike the environment-dependent behavior of $\hat{B}_{\rm{}pos}$ vs.\ $\hat{\nabla}_{\perp}I$ and $\vec{g}_{\rm{}pos}$ vs.\ $\hat{\nabla}_{\perp}I$. This persistent alignment is consistent with magnetically-guided accretion: sub-filaments channel material along field lines at several $10^{-3}$~$M_{\odot}$~yr$^{-1}$, sufficient to assemble the Ridge within $\sim{}10^6$~yr and sustain high-mass star formation.

This framework also explains why observed radial velocities ($\approx{}2~\rm{}km\,s^{-1}$) fall well below free-fall expectations ($\approx{}8~\rm{}km\,s^{-1}$): with the field nearly in the plane of the sky, only a small fraction of the accretion velocity projects along the line of sight.
\end{abstract}


\section{Introduction}\label{sec:intro}
Understanding what regulates star formation in molecular clouds remains a central question in astrophysics. The remarkably low efficiency of star formation, wherein only a few percent of molecular cloud mass converts to stars per free-fall time, suggests that processes beyond pure gravitational collapse govern the evolution of star-forming gas. Magnetic fields, alongside turbulence, have long been recognized as potentially playing a crucial role in regulating the gravitational collapse of star-forming gas \citep{McKee2007, crutcher2012:review}.

Early theoretical works on ambipolar diffusion and the mass-to-flux ratio as a key property governing the stability of magnetized clouds include \citet{Mestel1956} and \citet{mouschovias1976:magn_cl_i, mouschovias1976:magn_cl_ii}. See \citet{McKee1993:PP_III} for a review of these and other key concepts concerning magnetized clouds. Magnetically subcritical clouds are those with mass-to-flux ratios less than the critical value and cannot collapse until ambipolar diffusion or other processes weaken magnetic support, while supercritical clouds collapse on dynamical timescales.

Star formation unfolds within dense filamentary structures in molecular clouds \citep{Schneider1979, Hacar2023:PPVII_Filaments}. The \emph{Herschel} Space Observatory  revealed the ubiquitous presence of these elongated features \citep{Andre2010, Molinari2010}. The \emph{Herschel} Gould Belt Survey demonstrated that filaments are fundamental structures that host most of the dense gas from which stars form \citep{Arzoumanian2019}. Gravitationally unstable filaments fragment into chains of prestellar cores and protostars \citep{Andre2014, Hacar2023:PPVII_Filaments}. In both low- and high-mass star-forming regions, multiple filaments converge into hub-filament networks that funnel material into ridges where stars and clusters form \citep{Myers2009, Kumar2020, Andre2014, Hacar2023:PPVII_Filaments}.

Among the various agents that introduce anisotropy into the interstellar medium, including shocks, supernova feedback, and self-gravity, magnetic fields play a particularly pervasive role, imposing a preferred direction on gas dynamics across a wide range of scales. The Lorentz force preferentially resists gas motion perpendicular to magnetic field lines while allowing gas to flow freely along field lines \citep{Hennebelle2013, Li2014,Soler2017,Seifried2020:parallel-perpendicular}. This anisotropy has several observable consequences: gravitational contraction proceeds more readily along field lines than across them, creating flattened or elongated structures; turbulent motions become anisotropic in the presence of dynamically important fields \citep{Gonzalez-Casanova2017, Hu2019}; and the relative orientation between magnetic fields and density structures encodes information about the dynamical importance of the field \citep{Ibanez-Mejia2022, McGuiness2026}.

Magnetic fields in dense clouds are most readily traced via polarimetry of thermal dust emission: spinning elongated grains preferentially align with their angular momenta parallel to the ambient magnetic field, resulting in linearly polarized dust emission exhibiting a polarization angle perpendicular to the orientation of the magnetic field as described by the radiative aligned torques (RAT) model \citep{Dolginov1976, Lazarian2007, Andersson2015}. At the coarse $5\arcmin$ native resolution of \emph{Planck}, often smoothed further to $\sim{}10\arcmin$ to enhance signal-to-noise, a consistent picture has emerged across a range of scales: in diffuse regions, elongated density structures align parallel to the magnetic field, while in denser filaments the alignment tends to become perpendicular to the field \citep{Planck2016:XXXV_HRO}. In Taurus, \citet{Goldsmith2008} and \citet{Chapman2011} found that diffuse filaments with $A_V\approx{}1.25~\rm{}mag$ tend to be aligned with the magnetic field, while denser filaments are preferentially perpendicular, similar to the trends observed in the \emph{Planck} analysis. At higher angular resolution, \sofia/\hawc\ and JCMT/POL-2 studies reveal that dense filament ridges exhibit magnetic fields that are roughly orthogonal to the long axes \citep{Pattle2023:PPVII_MagneticFieldsCloudsCores, Stephens2025}. These multi-scale alignment patterns suggest that magnetic fields play a significant role in shaping filamentary clouds and channeling mass flows.

Despite this progress, several critical observational gaps remain. Observations at different scales often use different techniques and tracers, making it difficult to construct a coherent picture of magnetic field evolution from cloud scales ($\sim$10~pc) through filament scales ($\sim$0.1--1~pc) to core scales ($\sim$0.01~pc). High-resolution polarimetric observations at the characteristic $\sim$0.1~pc filament width scale, where the balances among turbulent, magnetic, and gravitational forces are expected to shift decisively towards gravity-dominated dynamics, are sparse. Therefore, direct constraints on how magnetic field strength evolves from diffuse envelopes to dense ridges within individual filaments are limited.

The \sofia/\hawc\ instrument \citep{Dowell2010,Harper2018} offered unique capabilities for addressing these gaps. Operating at stratospheric altitudes until \sofia's retirement in September 2022, \hawc\ largely avoided the significant atmospheric absorption that prevents ground-based far-infrared observations. 
Cold dust in star-forming regions (10--30~K) emits thermally
with a spectral peak near 100--300~$\mu$m; by observing at 214~$\mu$m,
close to this peak, \hawc\ was sensitive to the bulk of the cold dust column
and could detect polarized emission at lower column densities than
sub-millimeter ground-based facilities. This extended dynamic range is critical as it
allows polarization measurements to span from the diffuse cloud envelope
into the dense interior. Doing so captures the zones of potential transition from magnetic field
orientations at low column density to high column density. These are key diagnostics of the
interplay between gravity and magnetic support.
At a resolution of $\sim$18$''$, \hawc\ probed spatial scales of $\sim$0.05--0.2~pc for nearby clouds, directly matching the characteristic filament width identified by \emph{Herschel} \citep{Andre2014}, bridging the gap between large-scale \textit{Planck} observations and small-scale interferometric studies.

The \underline{S}tudy of \underline{I}nterstellar \underline{M}agnetic \underline{P}olarization: a \underline{L}egacy \underline{I}nvestigation of \underline{Fi}laments (SIMPLIFI) is a {\it SOFIA Legacy Program} (PI: T. Pillai) designed to exploit these capabilities. SIMPLIFI obtained 214~$\mu$m dust polarization observations of filamentary molecular clouds, probing the magnetic field morphology at spatial scales of around 0.1~pc. 
By targeting well-studied filamentary clouds spanning masses of 10--$2,000~M_{\odot}$, a range of evolutionary states, and star-forming activities, the SIMPLIFI project was designed to address how magnetic fields evolve from cloud scales down to core scales, and to ascertain the scales at which the fields become dynamically important. Key questions include whether magnetic fields set the mass accretion rate onto filaments, whether the fields are strong enough to influence fragmentation, and whether star formation efficiency is ultimately governed by magnetic field support. The overarching goal is to constrain the relative importance of magnetic fields, turbulence, and gravity in regulating star formation within filaments. In this first SIMPLIFI paper, we focus on DR21, one of the most massive and actively star-forming filaments in the sample; subsequent papers will present the full SIMPLIFI cloud sample and comparative analyses across diverse filamentary environments. 

\subsection{The DR21 Complex}\label{sec:dr21_intro}
The DR21 complex is located within the Cygnus~X molecular cloud complex, one of the most massive and active high-mass star-forming regions within 2~kpc of the Sun, containing numerous OB associations, Wolf-Rayet stars, \hii{} regions, massive protostars, and embedded clusters of young stellar objects \citep{Schneider2006, Kumar2007, Motte2007, Reipurth2008,Beerer2010}.
The mean distance to the Cygnus~X complex derived from trigonometric parallax measurements of methanol and water masers of multiple star-forming regions including DR21 is $1.40 \pm 0.08$~kpc \citep{Rygl2012}.

The DR21 complex contains a massive, elongated ridge — the DR21 Main Ridge — approximately 4~pc in length, oriented in the north-south direction. With a total mass $\approx{}2\times{}10^4\,M_{\odot}$ and column densities exceeding 10$^{23}$~cm$^{-2}$ (\citealt{schneider2010:dr21, hennemann2012:dr21}; Sec.~\ref{sec:segmentation}), it represents the densest and most massive filamentary structure in Cygnus~X. The Main Ridge is connected to lower-density sub-filaments that have been suggested to facilitate mass accretion into the central structure \citep{Kumar2007, schneider2010:dr21, hennemann2012:dr21, Hu2021, Ching2022:BISTRO-DR21}. The Main Ridge hosts multiple sites of active massive star formation, including the compact \hii{} region DR21 Main at the southern end, the protocluster DR21(OH) in the central region, and the W75S region to the north. Molecular line observations reveal complex kinematics consistent with large-scale gravitational collapse, with mass accretion rates of $\sim$10$^{-3}$~M$_\odot$~yr$^{-1}$ onto  DR21~Main 
and DR21(OH) \citep{schneider2010:dr21}. The Main Ridge also hosts prominent outflows \citep{Garden1991a, Garden1991b,Davis2007, Duarte-Cabral2013}. 

\subsection{Previous Magnetic Field Studies of DR21}
\label{sec:dr21_bfield}

The magnetic field in the DR21 complex has been studied across a wide range of spatial scales using diverse techniques over the past three decades, providing crucial context for our \sofia/\hawc\ observations.

\textit{Early submillimeter and infrared polarimetry}: The first submillimeter polarization measurements of DR21 revealed a ``polarization hole'' toward the dense core \citep{Minchin1994}, interpreted as evidence for complex magnetic field structure along the line of sight. Single-dish observations from 100~$\mu$m to 1.3~mm have consistently revealed a uniform plane-of-sky field perpendicular to the Main Ridge \citep{Kane1993, Minchin1994, Glenn1999, Vallee2006, Kirby2009}. \citet{Itoh1999} obtained H$_2$ 2.12~$\mu$m polarimetry of the DR21 outflow lobes, deriving a plane-of-sky field strength of $\sim 70\,\mu$G in the ambient medium. The 350~$\mu$m polarimetry surveys with the Hertz instrument at the Caltech Submillimeter Observatory \citep{Dotson2010} mapped DR21 Main, finding polarization patterns indicating an ordered field with a pronounced polarization hole toward the intensity peak. \citet{Kirby2009} used these data along with line width information to estimate a field strength of $\sim$2.5~mG in the DR21 Main region.

\textit{Zeeman observations}: DR21 Main and DR21(OH) have been important targets for Zeeman measurements across multiple tracers.  \citet{Roberts1997} used H~{\sc i} Zeeman observations to measure a line-of-sight field strength of $\sim$0.5~mG at the eastern edge of the DR21~Main 
outflow cavity. CN Zeeman observations indicate line-of-sight field strengths of 0.4--0.7~mG in DR21(OH) \citep{Crutcher1999, Falgarone2008}, while OH Zeeman observations measure $\sim$0.1~mG in the photodissociation region \citep{Koley2021}. \citet{Harvey-Smith2008} used \emph{MERLIN} 6.7~GHz methanol maser observations to infer the magnetic field geometry in DR21(OH). \citet{Momjian2017} detected the Zeeman effect in the 44~GHz Class I methanol maser toward DR21(OH), measuring field strengths of tens of milligauss in the dense gas traced by these masers at densities $n \sim 10^{7-8}$~cm$^{-3}$.

\textit{Large-scale Planck observations}: At large scales ($\sim$10\,pc), Planck 353\,GHz polarization observations reveal a fairly regular global magnetic field in the diffuse regions surrounding the DR21 complex, oriented at a position angle $\sim{}45\arcdeg$ (east of North) on the eastern side of the DR21 Main Ridge and parallel to the Galactic disk plane \citep{Ching2022:BISTRO-DR21}. This regular field becomes distorted toward the DR21 Main Ridge and smoothly connects to the  dense parsec-scale fields detected with JCMT/POL-2 (see below). The magnetic field orientation on the western side of DR21 Main Ridge has a complex structure unresolved by \emph{Planck}.

\textit{JCMT BISTRO survey}: At intermediate scales (0.1--1~pc), the JCMT BISTRO survey obtained 850~$\mu$m polarimetry of the DR21 complex with $\sim$14$''$ resolution \citep{Ching2022:BISTRO-DR21}. This study revealed ordered magnetic fields perpendicular to the parsec-scale Main Ridge of the complex, a configuration that is consistent with Magnetohydrodynamic (MHD) simulations of strongly magnetized media where material flows along field lines to accumulate in filamentary structures.  Using angular dispersion function analysis, \citet{Ching2022:BISTRO-DR21} derived plane-of-sky field strengths of 0.6--1.0~mG in the dense Main Ridge and showed that the mass-to-flux ratio is magnetically supercritical in the Main Ridge. 

\textit{Interferometric core-scale observations}: At small scales ($\sim$0.01--0.1~pc), interferometric polarimetry at sub-arcsecond resolution reveals more complex magnetic field morphologies in dense cores \citep{Lai2003, Girart2013, Ching2017}, suggesting that gravitational collapse and local dynamics distort the field at small scales \citep{Ching2018}. Submillimeter Array (SMA) 880~$\mu$m polarization observations targeted six massive dense cores within the DR21 Main Ridge \citep{Ching2017}. In stark contrast to the ordered field of the parsec-scale Main Ridge, the dust polarization reveals complex magnetic field structures within the cores. The major axes of the cores appear aligned either parallel or perpendicular to the filament-scale magnetic field, suggesting that the large-scale field played some role in core formation since no evidence for random orientations was found. However, within the cores themselves, the correlation between core elongation and core-scale magnetic field is weaker. The derived field strengths range from 0.4--1.7~mG. Virial analysis indicates that gravitational energy dominates over both magnetic and kinematic energies at filament scales, whereas kinematic energy becomes dominant over both magnetic and gravitational energies at core scales. For DR21(OH) specifically, \citet{Girart2013} found evidence for a toroidal field configuration within the core and a total field strength of 2.1~mG.

\subsection{This Work}
In this Paper~I of the SIMPLIFI series, we present new \sofia/\hawc\ $214~\rm{}\mu{}m$ observations of DR21, achieving a $20\farcs{}3$ full-width-at-half-maximum (FWHM) beam width ($\sim0.14~\rm{}pc$) and unprecedented sensitivity. 
As shown in the later comparison to JCMT/POL-2 data (Figure~\ref{fig:polcomp}), the SIMPLIFI \hawc\ observations detect polarized emission not only along the Main Ridge but throughout the network of sub-filaments and the ambient DR21 complex, enabling development of a continuous map of the plane-of-sky field morphology across $\sim$0.1--10~pc scales that could not be achieved from the ground. This dataset bridges the gap between \emph{Planck}’s large‐scale view, ground-based single-dish submm polarimetric studies, and mm/submm interferometric studies of cores, providing a homogeneous foundation for assessing magnetic influence on filamentary accretion and collapse.

Below, we describe the acquisition and reduction of the data (Sec.~\ref{sec:observations}) and introduce several techniques to characterize the structure of molecular clouds and their magnetic fields (Sec.~\ref{sec:methods}). We then apply these techniques to data on DR21 (Sec.~\ref{sec:data-analysis}) and interpret the results in the broader context of the region (Sec.~\ref{sec:interpretation}). Our study is summarized in Sec.~\ref{sec:summary}.

\begin{figure*}
\includegraphics[width=\linewidth]{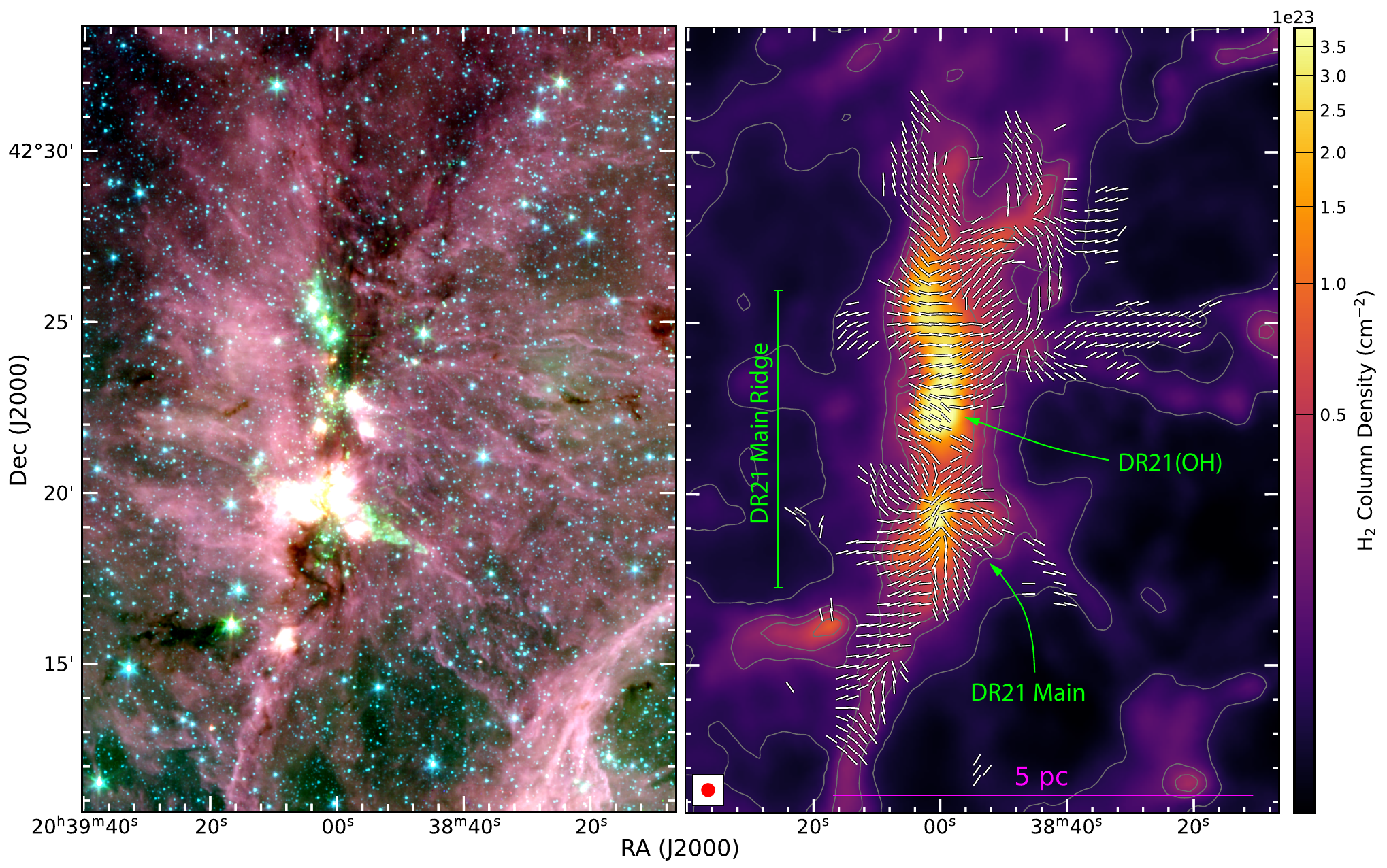}
\caption{\emph{Left:} \emph{Spitzer} three-color image of the DR21 complex (blue: IRAC1, green: IRAC2, red: IRAC4), rendered on 
the same pixel grid as the column density map in the 
right panel. \emph{Right:} Plane-of-sky magnetic field orientation inferred from \hawc\ 214\,$\mu$m polarization (rotated by $90\arcdeg$), 
overlaid on the \emph{Herschel}-derived H$_2$ column density 
map of \citet{pokhrel2020} (color scale). Contours mark H$_2$ 
column densities of 1, 2, 3, and 
$6\times{}10^{22}~\rm{}cm^{-2}$. Pseudovectors are spaced by 
$18\farcs{}2$ (4~pixels), approximately matching the 
$20\farcs{}3$ resolution of the smoothed \hawc\ data, which is also illustrated in the lower 
left corner.\label{fig:main}}
\end{figure*}

\section{\hawc\ Observations}\label{sec:observations}
\subsection{Observations}\label{sec:observations-HAWC+_observations}
The \hawc\ observations were acquired as part of \sofia\ project 09-0215 (T.~Pillai, PI). Two observing modes were employed. The chop-nod  (C2N) mode is the standard \sofia\ polarimetric observing mode and benefits from a mature, well-validated data reduction pipeline. On-the-fly (OTF)  mapping\footnote{This mode is called \texttt{OTFMAP} in the \emph{\sofia} documentation.} was offered on a ``shared-risk'' basis 
(Sec.~\ref{sec:hawc+-data-reduction}) and enables substantially more efficient  imaging of large fields, making it the practical choice for the SIMPLIFI  mosaic footprint. Accordingly, the first flight of this project (F776, 2021~Sep.~1) acquired data in both C2N and OTF modes over an overlapping field, allowing us to validate the OTF products against the established C2N reference (see Sec.~\ref{sec:hawc+-data-reduction}). All subsequent flights (F779, F780, F782, F784, F917, and F920, executed in September 2021 and 2022; 
\citealt{IPACSOFIADOI}) used OTF only. All data were acquired in \hawc\ band E, centered around a wavelength of $214~\rm{}\mu{}m$, which provides a nominal angular resolution of $18\farcs{}2$ (FWHM) \citep{Dowell2010, Harper2018}.

The C2N observations towards one position (named CygX\_mos1) were executed with a chop throw of $500\arcsec$, using a chop direction of $100\degr$ measured East of North. C2N used symmetric chopping, resulting in two reference positions against which the target field was measured. In this analysis, we only considered regions of the C2N target field that are at least a factor of five brighter than the reference fields, based on inspection of \emph{Herschel} 250\,$\mu$m maps, to reduce errors in the polarization angle resulting from polarized emission in the reference field (see Sec.~3.4 of \citealt{Chuss2019:OMC-1}). The OTF observations were centered on three locations (named CygX\_OTF, CygX\_DR21\_S, and CygX\_DR21\_S2) that, in combination, cover an irregularly bounded mosaic of about $18\arcmin\times{}22\arcmin$ shown in Fig.~\ref{fig:polcomp}~(left). The scan speed was $200\arcsec\,{\rm{}s^{-1}}$, and the scan duration was between 60~s and 120~s. Scanning followed a Lissajous pattern, in which the telescope 
traces a continuous two-dimensional trajectory with scan 
amplitudes varying between $60\arcsec$ and $200\arcsec$. A total of 171 scans were taken and processed.

Individual \sofia\ flights are documented in logs\footnote{\url{https://irsa.ipac.caltech.edu/data/SOFIA/docs/proposing-observing/flight-plans/index.html}}. We are not aware of substantial problems affecting the data used in this publication. To be specific, we sought to identify scans corrupted by issues with the \hawc\ cryogenic system, which occurred on some flights. Based on flight logs and \hawc\ sensor raw data (where, in our experience, images produced from the \texttt{SQ1Feedback} raw data field become noisy if the detector is warm), the specific data used in this publication are from flights not affected by cryogenic issues, or were taken before the cryogenic system warmed. We also screened raw data to remove observations without meaningful information on the precipitable water vapor (PWV) column, as these data are sometimes missing (i.e., PWV measurements of zero are recorded in the raw data). Such errors in PWV data can substantially affect the data calibration (see \citealt{Lopez-Rodriguez2022:SALSA_III-OTF} for a related but somewhat different case). None of the raw data used in this paper lack PWV information.

\subsection{Data Reduction\label{sec:hawc+-data-reduction}}
Science-ready C2N maps are a standard output of the \sofia\ data reduction pipeline. In the case of DR21, we used the standard output of the C2N processing pipeline as provided by the \sofia\ Archive at the Infrared Science Archive\footnote{\url{https://irsa.ipac.caltech.edu/data/SOFIA/HAWC_PLUS/L4/p12982/data/g6/F0776_HA_POL_09021544_HAWEHWPE_PMP_163-183.fits}}. Here, we used the C2N data as a reliable reference against which data obtained with the more novel OTF method could be compared. This is possible because C2N data reduction fundamentally builds on a comparison between the target and two reference fields. This comparison is straightforward and reliable, as long as the reference fields are free of emission (as in the case of DR21, Sec.~\ref{sec:observations-HAWC+_observations}).

The OTF observing mode used here was offered on a ``shared risk'' basis during the \sofia\ flights for this project. SIMPLIFI employed the OTF mapping technique as it permitted imaging much larger regions on the sky. The processing of OTF maps from \hawc\ was challenging as reference data were created within the target map through sophisticated filtering techniques and requires more care and attention than the processing of C2N observations. We used the \emph{\sofia} Redux package\footnote{We installed version 1.3.3 of the package from \url{https://github.com/SOFIA-USRA/sofia_redux}. Once launched in the interactive mode, the \texttt{Redux} window itself reports a version number of 2.14.14. Once \hawc\ data are loaded into the package, the window reports that version 3.2.1.dev0 of the \hawc\ Data Reduction Pipeline (DRP) has been activated.} to process the \emph{HAWC+} data. \texttt{Redux} converted the data from instrumental units to calibrated physical ones, filtered the data to remove artifacts, and gridded the data into maps. The package includes pipelines for the processing of \hawc\ data to a science-ready level. The performance of the \hawc\ pipeline has been validated in several publications \citep{Lopez-Rodriguez2022:SALSA_III-OTF, Li2022:Taurus-B211, Butterfield2024:FIREPLACE_I}.

We explored a broad range of \texttt{Redux} pipeline options on a dedicated workstation to arrive at the data reduction strategy used for the work presented here. All possible combinations between the parameters \texttt{rounds} (with values of 10, 20, 40, and 80), \texttt{smooth} (with values \texttt{beam} and \texttt{halfbeam}), \texttt{faint} (with values \texttt{True} and \texttt{False}), \texttt{extended} (with values \texttt{True} and \texttt{False}), \texttt{fixjumps} (with values \texttt{True} and \texttt{False}), and \texttt{grid} (with values \texttt{""} and 13) were examined, resulting in $4\cdot{}2^5=128$ separate reduction runs. We surveyed the parameter space to identify parameter sets that assure that (\textit{a})~the Stokes~$I$ signal from \emph{HAWC+} correlated well with the Stokes~$I$ signal from \emph{Herschel} at similar wavelengths, (\textit{b})~the polarized intensities from overlapping OTF and C2N maps are similar, (\textit{c})~the position angles of polarized emission from overlapping OTF and C2N maps were similar, and (\textit{d})~the polarized intensity was zero in regions where \emph{Herschel} detects no significant emission. A survey of the data reduction parameter space was needed for OTF processing as no standard strategy has been established by previous work. For example, some studies concluded that the OTF pipeline performs best when using just minimal iterative filtering (e.g., \citealt{Li2022:Taurus-B211} used very few iterative ``rounds'' and no other filter options in \texttt{Redux}; also see \citealt{Coude2026:DataReleaseFIELDMAPS}), while other work found that more substantial iterative filtering gave best results (e.g., \citealt{Butterfield2024:FIREPLACE_I} used many ``rounds'' and set additional options to process very large maps). It is plausible that these differences in data reduction strategies are driven by differences in map sizes, but detailed study is needed to establish the actual dependencies. Here, we used OTF data from the target CygX\_OTF for our analysis, as these maps overlap with the C2N observations.

Our experiments lead us to settle on a method utilizing a minimal number of iterative filter steps (i.e., $\texttt{rounds}=10$), half-beam smoothing ($\texttt{smooth}=\texttt{halfbeam}$), no filtering adapted for faint or extended emission ($\texttt{faint}=\texttt{False}$ and $\texttt{extended}=\texttt{False}$), no fixes in detector jumps ($\texttt{fixjumps}=\texttt{False}$), and no customized gridding ($\texttt{grid}=\texttt{""}$). We made these choices because they applied the least modifications to the raw data, but still delivered performance that was at least equal to those obtained when filtering and modifying the data in more substantial ways. Following the handbook, data acquired in band E have a 
diffraction-limited resolution of $18\farcs{}2$ (FWHM) and 
are gridded at $4\farcs{}55$ (i.e., Nyquist-sampled at 
one-quarter of the beam). The pipeline convolved the gridded 
data with a Gaussian kernel of 
$\mathrm{FWHM}=9\farcs{}1$ (half the beam width), yielding 
a restored map resolution of 
$(18\farcs{}2^2+9\farcs{}1^2)^{1/2}=20\farcs{}3$.

We had not attempted to optimize zero-level background subtraction \citep{Li2022:Taurus-B211}. This is in principle possible through the options of the \texttt{scanstokes} algorithm of the OTF data reduction pipeline. We refrained from doing this as the \hawc\ map contains sufficiently large areas free of emission. The agreement between OTF and C2N maps discussed below justified this approach a~posteriori.

To assess systematic effects in our data reduction, we compared position angles of polarized emission between the OTF and C2N maps. We restricted this comparison to measurements with  polarization signal-to-noise ratio $\ge 5$ (using uncertainties calculated by \texttt{Redux}), polarization fraction $\le 30\%$ (excluding nonphysical values; \citealt{PlanckCollaboration2020}), and position-angle uncertainty $\le 10\degr$. Under these criteria, the mean offset between OTF and C2N position angles was $5\degr$ with a standard deviation of $8\degr$. The small mean offset indicated that systematic errors between the two scan strategies were modest, while the $8\degr$ scatter was consistent with the per-measurement uncertainties reported by \texttt{Redux}, validating the formal error estimates.

\subsection{Data Products \label{sec:hawc+-data-products}}
The \hawc\ polarimetric products used here were generated by the tuned \texttt{Redux} pipeline as described in detail above. We summarize the standard definitions below to establish notation for the quantities used in our plotting and analysis. Stokes $I$ denotes the total intensity, while $Q$ and $U$ are components of the polarized intensity and may be written as
\begin{equation}
Q = P \cdot \cos(2\cdot{}\psi)
\quad \text{and} \quad
U = P \cdot \sin(2\cdot{}\psi) \, ,
\end{equation}
where $P$ is the polarized intensity and $\psi$ is the polarization position angle. In the IAU convention, $\psi$ is measured on the celestial sphere from North through East. From the Stokes parameters one defines the polarized intensity and observed polarization fraction as
\begin{equation}
P \equiv \sqrt{Q^{2}+U^{2}}
\quad \text{and} \quad
p_{\rm{}obs} \equiv \frac{P}{I} \, ,
\end{equation}
and a commonly used debiased estimator of the polarization fraction is
\begin{equation}
p \equiv \sqrt{p_{\rm obs}^{2}-\sigma_{p}^{2}}
\end{equation}
\citep{Wardle1974:PolarizationErrors, Simmons1985:PolarizationErrors, Vaillancourt2006:PolarizationConfidenceLimits}, where $\sigma_{p}$ is the uncertainty in $p_{\rm obs}$. The polarization angle is defined by
\begin{equation}
\psi \equiv \frac{1}{2}\arctan_2(U,Q),
\end{equation}
where $\arctan_2$ is the two-argument arctangent that returns the correct quadrant. In this paper, we adopted the pipeline-delivered values of $P$ and $p$, including bias corrections, and $\psi$ for plotting and subsequent analysis. We limited our analysis, as described below, to data passing the following quality cuts:
\begin{equation}
p / \sigma_p \ge 3 \, , \quad
p \le 0.3 \, , \quad {\rm{}and} \quad
I / \sigma_I \ge 50 \, ,
\end{equation}
where $\sigma_I$ is the uncertainty in Stokes~$I$. The constraint $p/\sigma_p \geq 3$ ensures a focus on reliable polarized emission. The limit $I/\sigma_I \geq 50$ is adopted as a high signal-to-noise threshold that balanced reliability against spatial coverage. The condition $p \leq 0.3$ is an empirical quality cut, motivated by 
\emph{Planck} measurements of $p_{\rm max} \approx 22\%$ in the diffuse ISM 
\citep{PlanckCollaboration2020}. In our map, only a single independent measurement exceeds this threshold, and it lies at low Stokes~$I$ where imperfect atmospheric subtraction 
would underestimate the true Stokes~$I$ and inflate $P/I$. These quality cuts were assumed in all that follows. These criteria left 519 independent polarization measurements, evaluated by the ratio between the map area and the area of the $20\farcs{}3$ beam of polarization measurements (using Eq.~\ref{eq:beam-area} below). Uncertainties in $I$, $Q$, and $U$ reported by Redux are below $0.016~\rm{}Jy\,pix^{-1}$ for 90\% of the mapped area.

The plane-of-sky magnetic-field orientation, $\hat{B}_{\rm pos}$, was obtained by rotating $\psi$ by $\pm90\arcdeg$. This interpretation assumes that aspherical dust grains align with their long axes preferentially perpendicular to the ambient magnetic field, as expected from radiative alignment torque (RAT) theory \citep{Lazarian2007}. In a separate paper of the SIMPLIFI series (Kumar et al. subm., Paper II), we present an extensive analysis of the DR21 polarization properties showing that RAT alignment remains efficient at high column densities, plausibly aided by internal illumination-driven alignment from embedded massive protoclusters. Accordingly, throughout this work we assumed that the $214~\mu{\rm m}$ polarization traces the magnetic-field morphology, averaged along the line of sight and as weighted by the column density of each emitting layer.

\subsection{Spatial Decomposition of DR21 Complex\label{sec:segmentation}}
The DR21 complex hosts several physically distinct components 
that must be distinguished before the polarimetric data can be 
analyzed. \emph{Spitzer} observations first detected several distinct filamentary structures converging on the DR21 Main Ridge, though they were not individually identified at the time \citep{Kumar2007}. Subsequent analysis by \citet{schneider2010:dr21} resolved two distinct sub-filaments within DR21, designated F1 and F3. Building on this work, \emph{Herschel} observations by \citet{hennemann2012:dr21} identified seven structures they described as streamers and that we label as sub-filaments here, including the two reported by \citet{schneider2010:dr21}. Listed from equatorial north to south in a clockwise direction, these are: N, F1N, F1S, F3N, F3S, SW, and S (see Fig.~\ref{fig:polcomp} for their on-sky locations). For brevity, we hereafter drop the ``F'' prefix where present. All of these features, with the exception of F1N which lies outside our mapped region, were detected in the \hawc\ Stokes~$I$ images at $214~\mu{}\rm{}m$ (see Section~\ref{sec:masks} below).

In addition to the sub-filaments, the DR21 Main Ridge hosts a powerful molecular outflow centered on DR21 Main \citep{Garden1991a, Garden1991b, Davis1996}. This outflow, one of the most energetic known in nearby star-forming regions, is expected to significantly impact the local gas dynamics and magnetic field structure and so is excluded from the analyses below to retain the focus on ascertaining the magnetic field roles (see Sec.~\ref{sec:masks} for details on the outflow mask).

\subsection{Overview of Magnetic Field Orientations}\label{sec:structure-overview}
The 214\,$\mu$m polarimetric data obtained with \hawc\ toward the DR21 complex are presented in Figure~\ref{fig:main}~(right). Polarization pseudo-vectors rotated by $90\degr$ are displayed with uniform length to emphasize the magnetic field orientation. The background color image shows the \emph{Herschel}-derived H$_2$ column density map from \citet{pokhrel2020}. The magnetic field orientation is predominantly east--west across the mapped region, roughly perpendicular to the north--south spine of the Main Ridge. In the sub-filaments, the field tends to follow the elongated structures, while in the Main Ridge it cuts across them. This contrast between the low- and high-density environments is the central pattern that motivates the quantitative analysis in Sec.~\ref{sec:data-analysis}.

\begin{figure*}
\centering
\includegraphics[width=\textwidth]{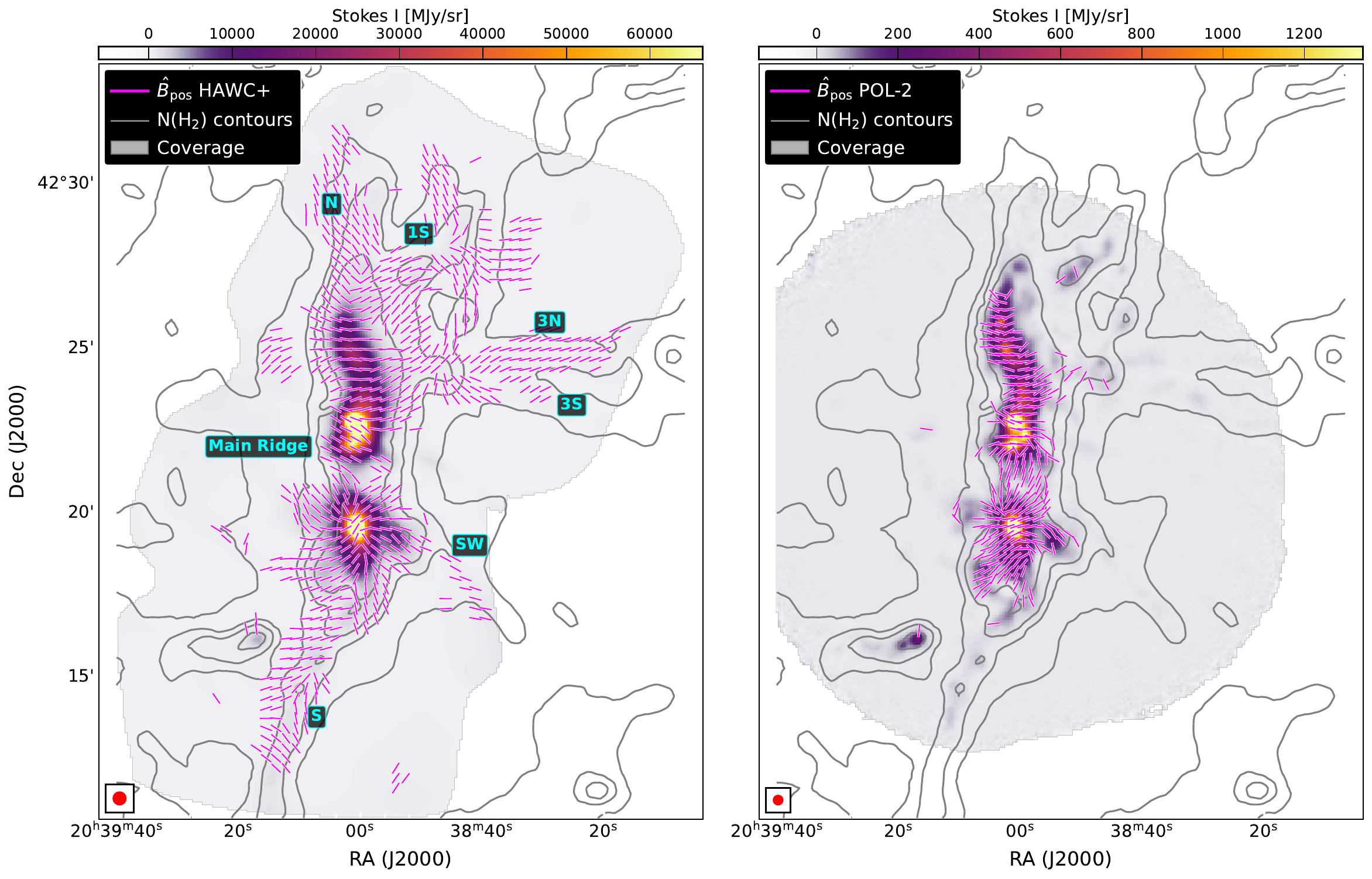}

\caption{Comparison of magnetic field orientation inferred from polarized dust emission observed with \sofia/\hawc\ at 214\,$\mu$m (left) and JCMT/POL-2 at 850\,$\mu$m (right) for $p/\sigma_{p} \geq 3.0$ \citep{Ching2022:BISTRO-DR21}. The background colorscale images show the respective Stokes $I$ intensities. Black contours in both panels indicate \emph{Herschel} H$_2$ column densities, as in Fig.~\ref{fig:main}.The pseudovectors show the plane-of-sky magnetic field orientation (\Bpos). Gray regions indicate extents of the respective sky mapping for polarization.  Sub-filament names from \citet{hennemann2012:dr21} are labeled on the HAWC+ panel. Filament F1N from that study lies outside the mapped region and is not shown.  Both panels are displayed on similar spatial scales to highlight differences in coverage and sensitivity between the two datasets.\label{fig:polcomp}}
\end{figure*}

As shown in Figure~\ref{fig:polcomp}, the \hawc\ observations detect polarized dust emission over a substantially more extended area than the most sensitive previous ground-based studies \citep{Ching2022:BISTRO-DR21}. Consistent with prior observations, the Stokes $I$ emission traces the DR21 Main Ridge, which is elongated in the north--south direction and hosts two well-known high-mass protoclusters: DR21(OH) near the center and DR21~Main to its south. North of DR21(OH), the magnetic field orientation is predominantly perpendicular to the Main Ridge spine, consistent with the findings of \citet{Ching2022:BISTRO-DR21}, with subtle indications of field curvature toward the dense interior of the Main Ridge. The field geometry becomes increasingly perturbed in the vicinity of DR21(OH) and DR21~Main, where the observed bent morphology is qualitatively consistent with flux-freezing models for magnetized, gravitationally contracting cores with spherical and prolate spheroidal geometries, respectively \citep{Myers2018, Myers2020}. Two arc-like polarization structures emanate from DR21~Main roughly along the northeast--southwest direction,  aligned with the powerful H$_2$ 
outflow that extends along a similar axis, visible as 
bright green diffuse emission in the \emph{Spitzer} RGB 
image (Fig.~\ref{fig:main}~left, where green traces the 
4.5\,$\mu$m IRAC2 band that includes H$_2$ line 
emission). This alignment suggests that the field morphology in this region is strongly influenced by outflow dynamics.

\citet{Ching2022:BISTRO-DR21} reported that the magnetic field orientation transitions from horizontal to northwest--southeast in the compact region bridging DR21(OH) and DR21~Main. Our observations, which have coarser angular resolution, marginally resolve this region across only 2--3 independent beams and yield no significant polarization detection there.

The \hawc\ observations reveal, for the first time, the magnetic field structure within the lower column density network of sub-filaments. The inferred field orientations in these structures connect smoothly to those of the Main Ridge. With the exception of sub-filament~S (see Section~\ref{sec:masks}), the magnetic fields in the sub-filaments are not orthogonal to their respective long axes. In the southern portion of sub-filament~S (south of
20:39:10 $+42$:14:00, J2000, see Section~\ref{sec:masks}), the plane-of-sky magnetic field orientation is predominantly perpendicular to the Main Ridge spine. Moving northward from this location, the field progressively reorients toward the Main Ridge axis, suggesting a deflection of the field.

Our sensitive Stokes $I$ and polarization maps also uncover a previously unidentified linear feature that crosses sub-filament~1S in projection (see Figure~\ref{fig:masks}) and extends roughly along a north--south axis. The magnetic field within this structure, which we designate the ``1S-Crossing'' (or ``1S-X''), is well-ordered and aligned parallel to its long axis, and hence orthogonal to sub-filament~1S, suggesting it may represent a dynamically distinct component, or a coherent accretion channel feeding into the Main Ridge.

Finally, we detect sparse polarization vectors toward the massive dense core N56 \citep{Motte2007} to the south, as well as along a filamentary feature extending eastward from DR21~Main. The latter corresponds to the SW sub-filament (labeled in Fig.~\ref{fig:polcomp}) identified in \emph{Herschel} observations \citep{hennemann2012:dr21} and detected also in JCMT/POL-2 data \citep{Ching2022:BISTRO-DR21}. Due to the limited number of independent \hawc\ polarization measurements, these peripheral features are not analyzed further in this work.

\section{Supporting Methods \label{sec:methods}}

\subsection{Analysis Masks and Filament Properties\label{sec:masks}}
\begin{figure}
\centering
\includegraphics[width=\linewidth]{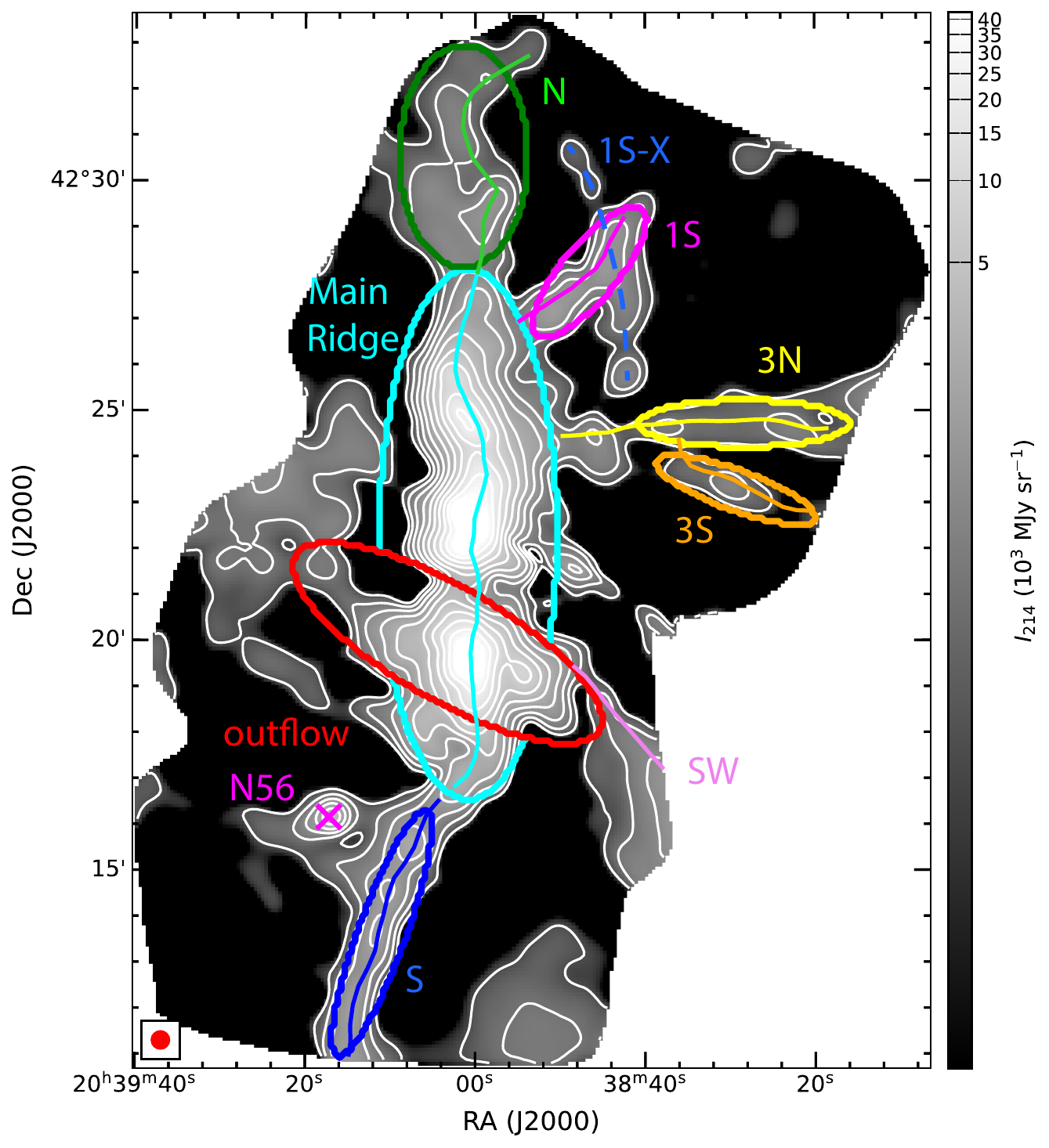}
\caption{\sofia/\hawc\ 214\,$\mu$m Stokes~$I$ image of DR21 in grayscale and contours with overlaid analysis masks (\emph{colored boundaries}) used to separate the Main Ridge from the system of sub-filaments. \emph{Thin colored lines} indicate the spines of the Main Ridge and sub-filaments, as derived from the \emph{Herschel}-based column density map. The \emph{red ellipse} marks the outflow region near DR21 Main, excluded from this study. A \emph{magenta cross} marks the approximate location of the dense cores N56. A \emph{dashed blue line} shows the location of the sub-filament 1S-X.\label{fig:masks}}
\end{figure}

Based on the physical components described in 
Sec.~\ref{sec:segmentation}, we define spatial masks 
for our analysis: (1) the DR21 Main Ridge, (2) the surrounding system of sub-filaments, and (3) the outflow region around DR21 Main. We excluded the latter from our analysis due to its exceptional energetics, as the intense conditions may introduce non-RAT alignment mechanisms (e.g., mechanical alignment via gas-grain drift or suprathermal spin-up, \citealt{Andersson2015}). Such effects would compromise the reliability of polarization as a tracer of magnetic field orientation, potentially biasing our $B$-field analysis. 
To generate the outflow mask, we used \emph{Spitzer IRAC} Band 2 data, which traces H$_2$ emission, ensuring complete coverage of the known H$_2$ outflow and extending well beyond the CO outflow emission \citep{Zapata2012}. The masks in general were defined as elliptical regions that encompass significant ($P/\sigma_P \ge 3$) polarized emission in each respective area while clearly separating them. These masks are shown in Figure~\ref{fig:masks}.

Here we trace the DR21 Main Ridge and the \citet{hennemann2012:dr21} sub-filaments on the basis of an $\rm{}H_2$ column density map derived by \citet{pokhrel2020} using \emph{Herschel} data. As indicated in Fig.~\ref{fig:masks}, the peak intensities are manually traced using a series of markers spaced by $<15\arcsec$, which is less than half the resolution of the \emph{Herschel} map. The mass and sky-projected length of these sub-filaments are characterized in Table~\ref{tab:subfilaments}. The mass for a given sub-filament was determined by summing the mass over all pixels separated by less than one $36\arcsec$ \emph{Herschel} FWHM beam-width from the central spine. This choice is motivated by the compact appearance of filaments, and uncertainties resulting from this choice are discussed below in this section. This is equivalent to adopting a characteristic sub-filament width $w=0.5\,\rm{}pc$ at the DR21 distance. Calculations outlined in Sec.~\ref{sec:guided-accretion_rates} are used to calculate $N_{\rm{}H_2}=(M/w^2)/\mu_{\rm{}H_2}$ and $n_{\rm{}H_2}=(M/w^2)/(\mu_{\rm{}H_2}\cdot{}\ell)$, estimate of the column and volume density one would obtain on lines of sight oriented \emph{along} the sub-filaments. We also determined the mass within 1~pc of the central line of the DR21 Main Ridge. We found a mass reservoir $M=1.7\times{}10^4\,M_{\odot}$ over a projected length $\ell=4.7~\rm{}pc$, giving a mass-to-length ratio of $M/\ell=3,650\,M_{\odot}\,\rm{}pc^{-1}$.

\emph{Herschel}'s angular resolution sets the lower limit to $w$ adopted above, and it is reasonable to consider larger values. For reference we experiment with doubling $w$ to 1.0~pc, equivalent to permitting angular offsets from filament spines of up to $72\arcsec$. This  increases $M$ and $M/\ell$ by a factor 1.5--2.0, while reducing $N_{\rm{}H_2}$ and $n_{\rm{}H_2}$ by a factor 2.0--2.5. This sensitivity to $w$ introduces some uncertainty in absolute filament properties. In practice the choice of $w$ corresponds physically to which section of the filament is being characterized: a small $w$ traces the dense inner spine (Table~\ref{tab:subfilaments}), while larger values of $w$ include more diffuse and extended envelope material. The values in Table~\ref{tab:subfilaments}, and the following sections using these values, should accordingly be read as focusing on the dense inner parts of the sub-filaments.

Observational uncertainties impose uncertainties at a level of several 10\% on column densities like those shown in Fig.~\ref{fig:main}~(right), which are derived from dust continuum emission \citep{Barnes2020:LEGO_ii}. Additional uncertainties by a factor $\sim{}2$ result from our limited knowledge of conversion factors between gas column densities and dust emission intensities \citep{kauffmann2010:mass-size-i}. Note, though, that these uncertainties are \emph{global}, i.e., errors will move all measurements of mass and column density into the same direction by the same factor. This means that mass and column density ratios are known to much higher precision. Properties such as the projected column density $N_{\rm{}H_2}=(M/w^2)/\mu_{\rm{}H_2}$, where $w$ is 
the sub-filament width (Sec.~\ref{sec:guided-accretion_rates}), 
are affected by additional conceptual uncertainties due 
to our limited knowledge of cloud structure. We did not 
formally track these uncertainties.

\begin{table}
\centering
\caption{Properties of sub-filaments (Secs.~\ref{sec:segmentation} and \ref{sec:guided-accretion_rates}).\label{tab:subfilaments}}
\begin{tabular}{clllll}
\hline\hline
Sub-filament & $M$ & $\ell$ & $M/\ell$ & $N_{\rm{}H_2}^{\dagger}$ & $n_{\rm{}H_2}^{\ddagger}$\\
 & $M_{\odot}$ & pc & $M_{\odot}\,\rm{}pc^{-1}$ & $10^{23}~\mathrm{cm^{-2}}$ & $10^4~\mathrm{cm^{-3}}$\\
\hline
N & 798 & 2.4 & 338 & 1.4 & 2.0 \\
1S & 760 & 1.4 & 556 & 1.4 & 3.2\\
3N & 503 & 2.4 & 210 & 0.9 & 1.2\\
3S & 322 & 2.0 & 164 & 0.6 & 0.9\\
SW & 454 & 2.3 & 197 & 0.8 & 1.1\\
S & 980 & 3.0 & 332 & 1.7 & 1.9\\
\hline
\multicolumn{6}{l}{$^{\dagger}$calculated as $N_{\rm{}H_2}=(M/w^2)/\mu_{\rm{}H_2}$}\\
\multicolumn{6}{l}{$^{\ddagger}$calculated as $n_{\rm{}H_2}=(M/w^2)/(\mu_{\rm{}H_2}\cdot{}\ell)$}
\end{tabular}
\end{table}

\subsection{Plane-of-Sky Gravitational Acceleration\label{sec:G_pos}}
\begin{figure*}
\centering
\includegraphics[width=\textwidth]{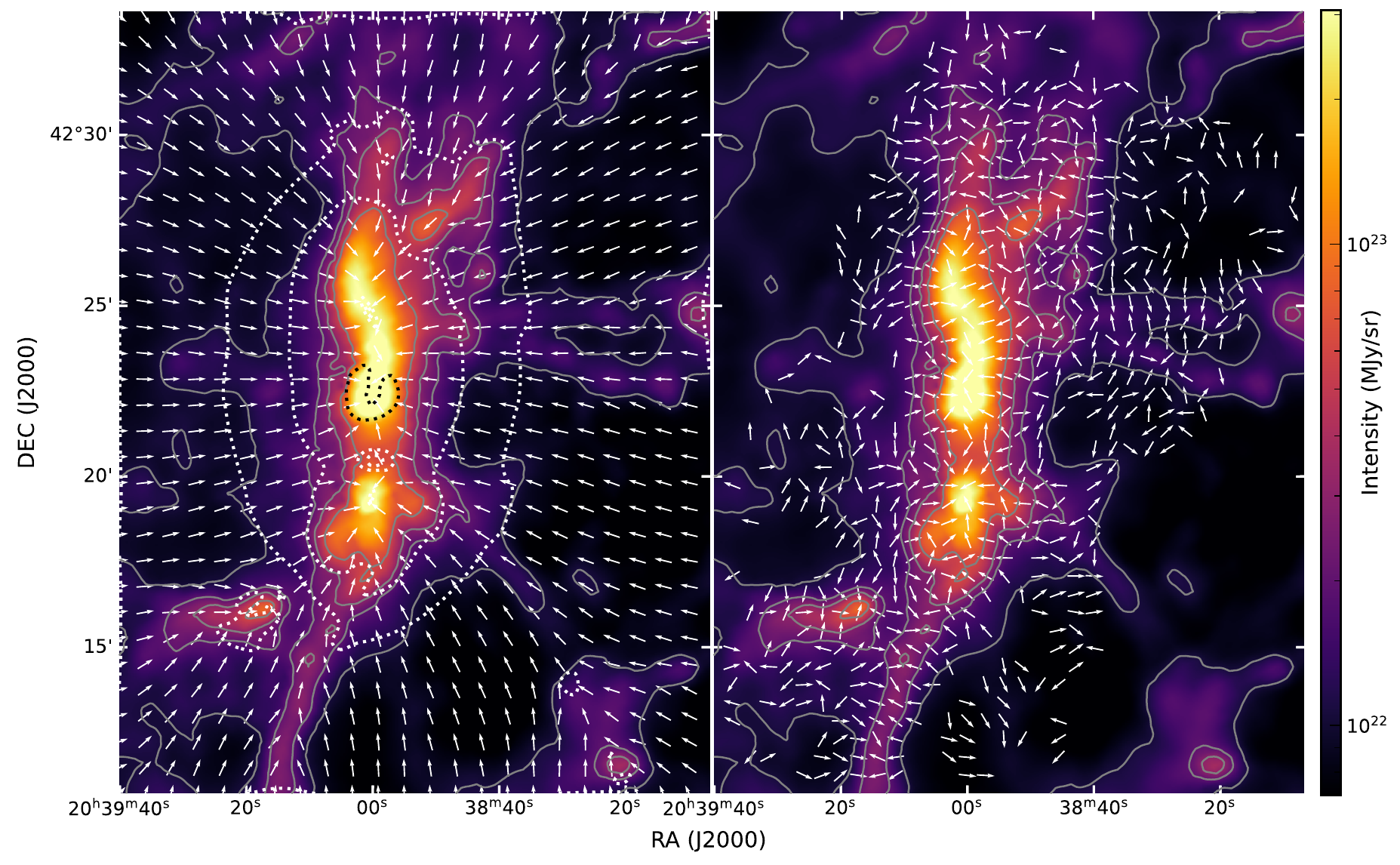}
\caption{The \emph{left panel} depicts the direction of the gravitational acceleration field $\vec{g}_{\rm{}pos}$ in the form of white vectors. The magnitude of $\vec{g}_{\rm{}pos}$ is indicated by dotted contours drawn at accelerations of $[0.5,1,5,10,50]\cdot{}10^{-10}~\rm{}m\,s^{-2}$. The \emph{right panel} depicts the direction of intensity gradients $\vec{\nabla}I$ obtained from \emph{HAWC+} data, drawn in locations where the uncertainty in the position angle of $\vec{\nabla}I$ is $\le{}15\degr$. Vectors for $\vec{g}_{\rm{}pos}$ are spaced by 46\arcsec. Vectors for $\vec{\nabla}I$ are spaced by 32~arcsec. Gray contours and the false color background are taken from Fig.~\ref{fig:main}~(right) and indicate \emph{Herschel} H$_2$ column densities.
\label{fig:gradients}}
\end{figure*}

The magnitude and direction of gravitational acceleration is relevant for our understanding of the dynamics in the DR21 complex. We derive an idealized estimate of the gravitational acceleration by placing all mass in a sheet parallel to the plane of the sky and located at the DR21 distance, as for example previously assumed by \citet{Koch2012:IntensityGradients}. In that case, the gravitational acceleration at location $\vec{x}_0$ is
\begin{equation}
\vec{g}_{\rm{}pos}(\vec{x}_0) =
    G \int_A \Sigma(\vec{x}) \cdot 
    \frac{
        \vec{x}-\vec{x}_0
        }{
        \left|\vec{x}-\vec{x}_0\right|^3
        } \, {\rm{}d}^2x \, ,
    \label{eq:estimate-g}
\end{equation}
where $\Sigma(\vec{x})$ is the mass surface density, $G$ is the gravitational constant, and $\vec{x}$ gives physical positions in the plane of the sky (e.g., measured in meters). $A$ indicates the area over which the integration is performed, which in this study we take to be the entire region shown in Fig.~\ref{fig:main}. The resulting gravitational vector field is illustrated in Fig.~\ref{fig:gradients}~(left). The vectors 
converge onto the DR21 Main Ridge, reflecting its 
dominance as the primary mass concentration in the 
region. Along the ridge itself, the gravitational field 
is directed largely along its length, potentially funneling material 
toward the central hub around DR21~Main.

Systematic errors in distance measurements (which would bias $\vec{x}$ by a constant factor) and conversion to column density (which would bias $\Sigma[\vec{x}]$ by a constant factor) affect the magnitudes of $\vec{g}_{\rm{}pos}$, but not their directions.

\subsection{Intensity Gradients as a Measure of Spatial Cloud Structure}
Some aspects of our analysis require a characterization of the spatial structure of the cloud. The orientation of intensity gradients has become a useful metric for that purpose \citep{Soler2013:imprint-magnetic-field}. We calculate gradients $\vec{\nabla}I$ for the Stokes~$I$ component of the \emph{HAWC+} map using the \texttt{gradient} function from \texttt{NumPy} \citep{Harris2020:NumPy}. This operation is performed at the original $4\farcs{}55$ gridding of the data. The output from \texttt{gradient} is used to calculate the position angle of the intensity gradient. The uncertainty of the position angle in a given pixel is calculated via a Monte~Carlo approach. First, a series of 50 synthetic intensity maps is generated by adding Gaussian noise with a standard deviation equal to the observational uncertainties to each pixel in the observed intensity map. Second, intensity gradient position angles are measured in the synthetic intensity maps, and the standard deviation in these synthetic angles is calculated for each pixel after unwrapping their orientations. This standard deviation is taken to be the uncertainty in orientations. The resulting vector field is illustrated in Fig.~\ref{fig:gradients}~(right).

The analysis below makes use of $\vec{\nabla}I$ rotated by $\pm{}90\degr$, to which we refer as $\hat{\nabla}_{\perp}I$. The hat accent indicates that this property is a pseudovector that has a directional ambiguity of $\pm{}180\degr$, as a line orthogonal to a vector has no clear direction. This is an interesting property, as it is parallel to intensity contours that can be used to trace the orientation of elongated cloud density structure \citep{Koch2012:IntensityGradients, Planck2016:XXXV_HRO}.

One might intuitively expect $\vec{\nabla}I$ to be well aligned with $\vec{g}_{\rm{}pos}$. Such a trend would be plausible as $\vec{\nabla}I$ is directly sensitive to gradients in mass distribution, which in turn determine the gradients in the gravitational potential indicated by $\vec{g}_{\rm{}pos}$. However, as illustrated in Figure~\ref{fig:gradients} and Sec.~\ref{sec:G-vs-density}, the relative orientation between $\vec{\nabla}I$ and $\vec{g}_{\rm{}pos}$ is complex. This results from the integral in Eq.~(\ref{eq:estimate-g}), which makes $\vec{g}_{\rm{}pos}$ a property influenced by distant mass reservoirs, unlike $\vec{\nabla}I$ which is purely driven by nearby pixels.

\subsection{Characterization of the Alignment between Vector Fields\label{sec:methods-alignment-measurement}}
Much of the analysis in Sec.~\ref{sec:data-analysis} focuses on the relative alignment between vectors and pseudovectors. Studies of \emph{Planck} data by, e.g., \citet{jow2018} and \citet{soler2019} have established the projected Rayleigh statistic (PRS) as a tool that quantifies the alignment between pseudovectors. The PRS is derived from the classic Rayleigh test used in circular statistics, but modified to handle axial data, where angles $\Phi$ and $\Phi{}\pm{}180\degr$ are considered equivalent \citep{jow2018}. The PRS is represented by the quantity Z, defined as
\begin{equation}
\begin{split}
Z
    &= \frac{\sum_i^{n_{\rm{}ind}} \cos(2\cdot{}\Phi_i)}{(n_{\rm{}ind}/2)^{1/2}} \\
    &= (2 \cdot n_{\rm{}ind})^{1/2} \cdot
    \langle \cos(2\cdot{}\Phi_i) \rangle \, ,
\label{eq:PRS-def}
\end{split}
\end{equation}
where $\Phi_i$ are the relative orientation angles, $n_{\rm{}ind}$ is the number of independent data samples, and $\langle\cos(2\cdot\Phi_i)\rangle=[\sum_i^{n_{\rm{}ind}} \cos(2\cdot{}\Phi_i)]/n_{\rm{}ind}$ is the mean value of $\cos(2\cdot\Phi_i)$. The PRS is designed to determine whether the observed set of relative orientations $\Phi_i$ indicate preferential parallel alignment corresponding to $|\Phi_i|<45\degr$, which gives $Z>0$, or a perpendicular orientation corresponding to $|\Phi_i|>45\degr$, which gives $Z<0$. \citet{fissel2019} highlight that the angles $\Phi_i$ are typically measured on a per-pixel basis, and that the pixel number $n_{\rm{}pix}$ can greatly exceed $n_{\rm{}ind}$. They show that
\begin{equation}
Z =
    \left( \frac{n_{\rm{}ind}}{n_{\rm{}pix}} \right)^{1/2} \cdot
    \frac{\sum_i^{n_{\rm{}pix}} \cos(2\cdot{}\Phi_i)}{(n_{\rm{}pix}/2)^{1/2}}
\label{eq:definition-Z}
\end{equation}
can be used to determine $Z$ from a summation over pixels. The on-the-fly (OTF) data reduction done for \emph{HAWC+} data involves restoration of independent\footnote{OTF data dumps from locations separated by less than a telescope beam constitute independent 
measurements, as each carries an independent noise 
realization. It is the gridding kernel, not the beam 
itself, that introduces correlations between 
neighboring map pixels and thus sets 
$n_{\rm pix}/n_{\rm ind}$ via 
Eq.~(\ref{eq:pixels-per-kernel}). In the standard 
\emph{HAWC+} setup the kernel scales with the beam, 
but the two are conceptually distinct. We further note 
that astrophysical correlations between adjacent lines 
of sight are captured separately by $\sigma_{\rm gen}$ 
(Sec.~\ref{sec:methods-alignment-measurement}).} measurements with a Gaussian kernel of area $\Omega_{\rm{}kernel}$. The size of this gridding kernel determines the scale at which pixels become uncorrelated, so that
\begin{equation}
n_{\rm{}ind} / n_{\rm{}pix} =
    \Omega_{\rm{}pix} / \Omega_{\rm{}kernel} \, .
\label{eq:pixels-per-kernel}
\end{equation}
The solid angle subtended by a square-shaped pixel with edge length $x_{\rm{}pix}$ is $\Omega_{\rm{}pix}=x_{\rm{}pix}^2$. The solid angle subtended by the restoring kernel with FWHM of $\vartheta_{\rm{}kernel}$ is \citep{kauffmann2008:mambo-spitzer}
\begin{equation}
\begin{split}
\Omega_{\rm{}kernel}
    &=
    \frac{\pi}{4\cdot\ln(2)} \cdot \vartheta_{\rm{}kernel}^2 \\
    &\approx
    1.13 \cdot \vartheta_{\rm{}kernel}^2 \, .
\label{eq:beam-area}
\end{split}
\end{equation}
SIMPLIFI uses standard gridding and half-beam smoothing as defined in the default \emph{HAWC+} data processing setup. In this case $x_{\rm{}pix}=\vartheta_{\rm{}beam}/4$ and $\vartheta_{\rm{}kernel}=\vartheta_{\rm{}beam}/2$, so that $n_{\rm{}ind}=[\ln(2)/\pi]\cdot{}n_{\rm{}pix}\approx{}0.22\cdot{}n_{\rm{}pix}$. Application of this scheme to the map parameters from \citet{fissel2019} reproduces their relation between $n_{\rm{}ind}$ and $n_{\rm{}pix}$.

\citet{jow2018} demonstrate that the standard deviation of $Z$ is $\sigma_{\rm{}flat}(Z)=1$ if the $\Phi_i$ are drawn from a flat random distribution spanning all angles. They also show that the standard deviation of $Z$ is

\begin{equation}
\begin{split}
\sigma_{\rm{}gen}(Z)
    &= \left( \frac{ 2 \cdot \sum_i^{n_{\rm{}ind}} \cos^2(2 \cdot \Phi_i)}{n_{\rm{}ind}} - \frac{Z^2}{n_{\rm{}ind}} \right)^{1/2} \\
    &= \left( 2 \cdot \left\langle \cos^2(2 \cdot \Phi_i) \right\rangle - \frac{Z^2}{n_{\rm{}ind}} \right)^{1/2} \\
    &= \left( 2 \cdot \left\langle \cos^2(2 \cdot \Phi_i) \right\rangle - 2 \cdot \left\langle \cos(2 \cdot \Phi_i) \right\rangle^2 \right)^{1/2}
\label{eq:Z-unc-common_angle}
\end{split}
\end{equation}
for generalized distributions of $\Phi_i$, where the last step uses Eq.~(\ref{eq:PRS-def}). The impact of individual observational uncertainties $\sigma(\Phi_i)$ can be estimated via Gaussian error propagation (Eq.~[\ref{app-eq:Z-unc-general}]),
\begin{equation}
\begin{split}
\sigma_{\rm{}obs}(Z)
    &=
    \left[
        \sum_i^{n_{\rm{}ind}}
            \left(
                \frac{\partial Z}{\partial \Phi_i} \cdot \sigma(\Phi_i)
            \right)^2
    \right]^{1/2} \\
    &=
    \left(
        \frac{8}{n_{\rm{}ind}} \cdot
        \sum_i^{n_{\rm{}ind}}
            \left(
                \sin(2\cdot{}\Phi_i) \cdot \sigma(\Phi_i)
            \right)^2
    \right)^{1/2} \\
    &=
    8^{1/2} \cdot
        \left\langle
            \sin^2(2\cdot{}\Phi_i) \cdot \sigma^2(\Phi_i)
        \right\rangle^{1/2}
    \, .
\end{split}
\label{eq:uncertainty_Z-individual}
\end{equation}
These uncertainty measures can be used in the interpretation of $Z$. Given $Z/\sigma_{\rm{}flat}(Z)=Z$, values $|Z|>3$ are (following \citeauthor{jow2018}) generally interpreted as a very strong signal for parallel or perpendicular orientation of investigated properties. In practice, though, distributions of $\Phi_i$ in a region of interest are more concentrated than the flat distribution assumed to evaluate $\sigma_{\rm{}flat}(Z)$. This makes $|Z|/\sigma_{\rm{}gen}(Z)$ a better metric to assess whether, compared to the dispersion among observed angles $\Phi_i$, the significance of the mean angle $\langle\cos(2\cdot\Phi_i)\rangle$ is high enough to allow a meaningful interpretation.  In this publication we therefore use a threshold $|Z|/\sigma_{\rm{}gen}(Z)>3$ to establish the presence of a very strong signal for parallel or perpendicular orientation.

The significance of $\langle\cos(2\cdot\Phi_i)\rangle$ with respect to observational uncertainties is given by $|Z|/\sigma_{\rm{}obs}(Z)$. The key conceptual difference between the latter two measures of significance is that $|Z|/\sigma_{\rm{}gen}(Z)$ takes variations in $\Phi_i$ driven by both source structure and observational uncertainties into account. This can for example be used to assess whether $\langle\cos(2\cdot\Phi_i)\rangle$ can be constrained meaningfully in a region in which the orientation of $\hat{B}_{\rm{}pos}$ changes due to astrophysical processes. The metric $|Z|/\sigma_{\rm{}obs}(Z)$, by contrast, assesses whether another independent measurement of $Z$ would deliver the same result.

{Equation~(\ref{eq:PRS-def}) shows that $Z$ directly depends on $n_{\rm{}ind}$, i.e., the value of $Z$ is not an immediate indicator of the sample of angles $\Phi_i$, as for example characterized through the average $\langle{}\cos(2\cdot{}\Phi_i)\rangle$. This is a deliberate aspect of $Z$, which was designed to be a metric of signal strength, not of orientation \citep{jow2018}. We therefore also use projected pixel-by-pixel alignment components
\begin{equation}
z_i = \cos(2\cdot{}\Phi_i)
\label{eq:zi-def}
\end{equation}
in our analysis, as these are more closely related to the properties of the $\Phi_i$ samples. Similar to $\sigma_{\rm{}gen}(Z)$, uncertainties
\begin{equation}
\sigma_{\rm{}gen}(\langle{}z_i\rangle) = 
    \sigma(z_i) / n_{\rm{}ind}^{1/2}
\label{eq:unc-mean-z_i}
\end{equation}
describe deviations from the averages $\langle{}z_i\rangle$ resulting from observational uncertainties and intrinsic source structure.}

An alternative, more conservative approach in these calculations would be to define the number of independent measurements based on the telescope beam area $\Omega_{\rm{}beam}$, so that $n_{\rm{}ind}/n_{\rm{}pix}=\Omega_{\rm{}pix}/\Omega_{\rm{}beam}$. Substitution for our specific case in the equations above shows that doing so would reduce $n_{\rm{}ind}$ by a factor 4, $|Z|$ by a factor 2, and leave $\sigma_{\rm{}gen}(Z)$ and $\sigma_{\rm{}obs}(Z)$ unchanged. Following \citet{fissel2019}, published applications of the PRS to far-infrared polarimetry have consistently employed the gridding kernel to define the correlation scale, and so we adopt the same convention here. The more conservative approach, with $|Z|$ reduced by a factor 2 does not alter the discussion below due to the high signal-to-noise ratio in the data points driving our discussion.

\begin{figure*}
\centering
\includegraphics[width=\textwidth]{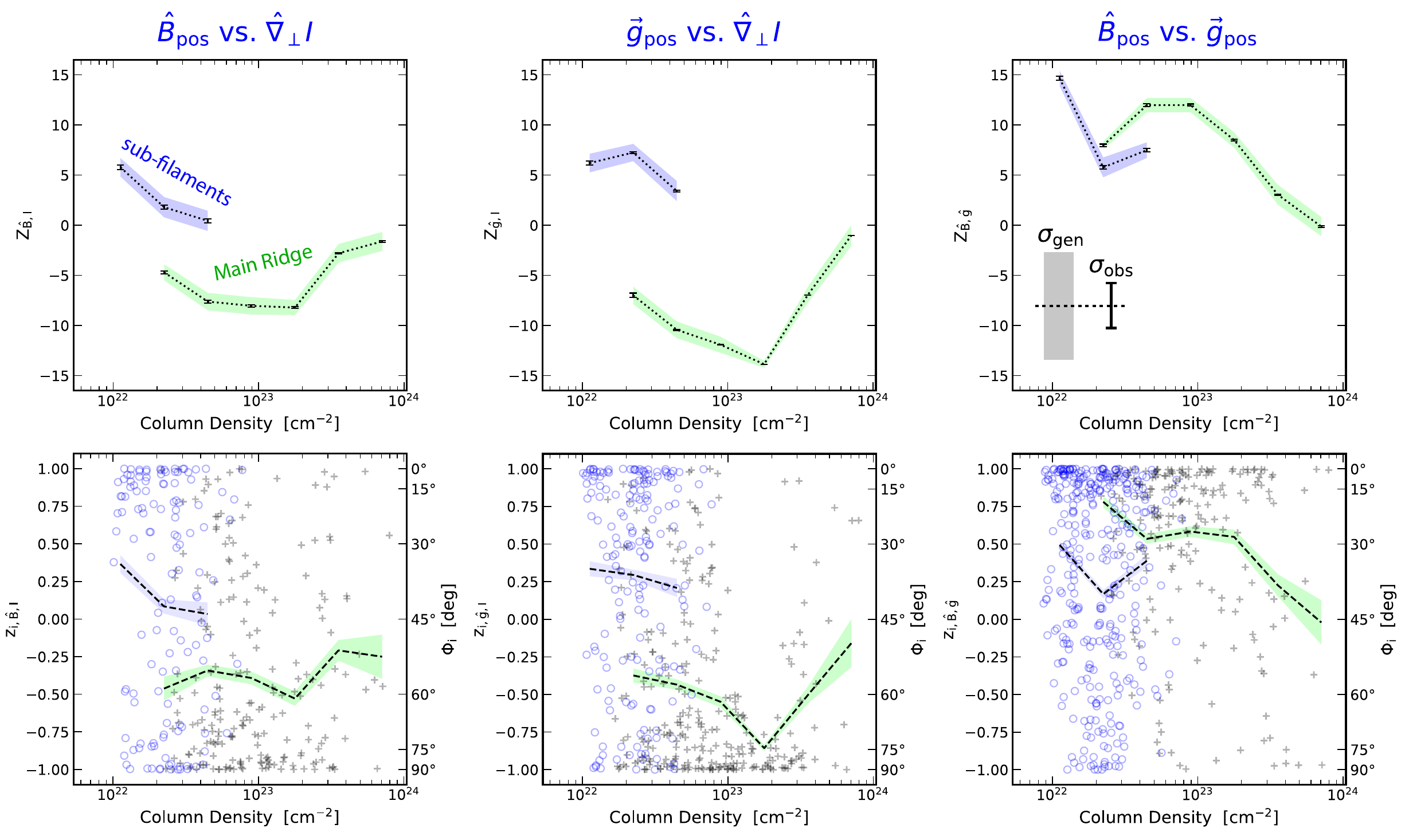}

\caption{Projected Rayleigh statistic $Z$ (\emph{top row}; 
Eq.~[\ref{eq:PRS-def}]) and per-pixel alignment components $z_i$ 
(\emph{bottom row}; Eq.~[\ref{eq:zi-def}]) as a function of column 
density. \emph{Blue shading/empty circles}: sub-filaments; 
\emph{green shading/gray crosses}: DR21 Main Ridge. Columns show the 
relative orientation between magnetic field and intensity gradient 
(\emph{left}; Sec.~\ref{sec:B-vs-density}), self-gravity and 
intensity gradient (\emph{middle}; Sec.~\ref{sec:G-vs-density}), 
and magnetic field and self-gravity (\emph{right}; 
Sec.~\ref{sec:BvsG}). For $Z$, \emph{error bars} give 
$\sigma_{\rm obs}(Z)$ (Eq.~\ref{eq:uncertainty_Z-individual}), 
reflecting individual measurement uncertainties, while 
\emph{shading} gives $\sigma_{\rm gen}(Z)$ 
(Eq.~\ref{eq:Z-unc-common_angle}), our preferred uncertainty 
measure that also captures internal structural diversity 
(Sec.~\ref{sec:methods-alignment-measurement}). An \emph{inset 
legend} in the upper-right panel illustrates both uncertainty types. 
For $z_i$, \emph{dashed lines} connect bin-averaged $\langle z_i 
\rangle$ values and \emph{shading} indicates 
$\sigma_{\rm gen}(\langle z_i 
\rangle)$.\label{fig:relative-orientations_BGI}}

\end{figure*}

\section{Data Analysis\label{sec:data-analysis}}
We characterize the relative orientations among three key projected quantities: the magnetic field orientation $\hat{B}_{\rm{}pos}$, the direction of gravitational acceleration $\vec{g}_{\rm{}pos}$, and the orientation of intensity gradients after rotation by $\pm{}90\degr$, $\hat{\nabla}_{\perp}I$. In the following subsections, we examine each pair in turn --- first $\hat{B}_{\rm{}pos}$ vs.\ $\hat{\nabla}_{\perp}I$ (Sec.~\ref{sec:B-vs-density}), then $\vec{g}_{\rm{}pos}$ vs.\ $\hat{\nabla}_{\perp}I$ (Sec.~\ref{sec:G-vs-density}), and finally $\hat{B}_{\rm{}pos}$ vs.\ $\vec{g}_{\rm{}pos}$ (Sec.~\ref{sec:BvsG}) --- using data sorted into column density bins spanning a factor $10^{0.3}\approx{}2$ in $N({\rm{}H_2})$ and grouped by the cloud environments defined in Sec.~\ref{sec:segmentation}.

\subsection{Relative Alignment between Magnetic Field and Cloud Density Structure\label{sec:B-vs-density}}
The relative orientation between $\hat{B}_{\rm{}pos}$ and $\hat{\nabla}_{\perp}I$ is characterized in Figure~\ref{fig:relative-orientations_BGI}~(left), as expressed through the PRS-derived properties $Z_{\hat{B},I}$ and $\langle{}z_{i,\hat{B},I}\rangle$. In the following we use $\sigma(Z_{\hat{B},I})$ and $\sigma(\langle{}z_{i,\hat{B},I}\rangle{})$ as our preferred measures of uncertainty, as described in Sec.~\ref{sec:methods-alignment-measurement}.

We find that generally $Z_{\hat{B},I}/\sigma_{\rm{}gen}(Z_{\hat{B},I})>1$ in sub-filaments (Fig.~\ref{fig:relative-orientations_BGI}~[left, top]). This is equivalent to $\hat{B}_{\rm{}pos}$ being parallel to column density contours,  with $Z_{\hat{B},I}/\sigma_{\rm{}gen}(Z_{\hat{B},I})>3$ required for high significance. Assuming that cloud structure is typically elongated and filamentary \citep{Hacar2023:PPVII_Filaments}, this finding indicates that $\hat{B}_{\rm{}pos}$ is broadly aligned with elongated cloud structure in the sub-filaments. By contrast we find that $Z_{\hat{B},I}/\sigma_{\rm{}gen}(Z_{\hat{B},I})<-1$ in the Main Ridge, with highly significant values $Z_{\hat{B},I}/\sigma_{\rm{}gen}(Z_{\hat{B},I})<-3$ in many column density bins. The value of $|Z_{\hat{B},I}|$ decreases to relatively small values at low and high column density in the Main Ridge  (Fig.~\ref{fig:relative-orientations_BGI}~[left, top]). Inspection of the averages $\langle{}z_{i,\hat{B},I}\rangle$ indicates that this does not result from a strong change in the relative alignment $\Phi_i$ between $\hat{B}_{\rm{}pos}$ and intensity gradients (Fig.~\ref{fig:relative-orientations_BGI}~[left, bottom]), but rather reflects the dependence of $Z_{\hat{B},I}$ on the number of data points, $|Z_{\hat{B},I}|\propto{}n_{\rm{}ind}^{1/2}$ (Eq.~\ref{eq:PRS-def}). More generally we find that, at fixed column density, the $\langle{}z_{i,\hat{B},I}\rangle$ in the sub-filaments differ from those in the Main Ridge by much more than their respective uncertainties $\sigma_{\rm{}gen}(\langle{}z_{i,\hat{B},I}\rangle)$.

Measurements of $\langle{}z_{i,\hat{B},I}\rangle$ and $\sigma_{\rm{}gen}(\langle{}z_{i,\hat{B},I}\rangle)$ thus indicate that the relative alignment between $\hat{B}_{\rm{}pos}$ and $\hat{\nabla}_{\perp}I$ differs substantially between sub-filaments and the Main Ridge, even within a given column density bin. Cloud-scale trends in $Z_{\hat{B},I}$ and $z_{i,\hat{B},I}$, as determined when using HAWC+ data from all cloud regions except for outflow-affected areas, fall between those seen for the sub-filaments and the Main Ridge (Appendix~\ref{sec-app:B-vs-density_all}).

In summary, elongated cloud structure is primarily aligned with the magnetic field in the sub-filaments of relatively low column density, and perpendicular to the Main Ridge with a relatively high column density. This is consistent with the transition from preferentially parallel relative orientation at low column densities to preferentially perpendicular relative orientation at high column densities \citep{Planck2016:XXXV_HRO, soler2019}.

The data are consistent with a cloud-scale transition from cloud structure aligned with $\hat{B}_{\rm{}pos}$ to cloud structure perpendicular to $\hat{B}_{\rm{}pos}$ at a column density $N({\rm{}H_2})\sim{}10^{22}~\rm{}cm^{-2}$. Specifically, the cloud-scale evaluation of the PRS shows that $Z_{\hat{B},I}=0$ coincides with $N({\rm{}H_2})=2\times{}10^{22}~\rm{}cm^{-2}$ (Fig.~\ref{fig-app:IvsB-all}), we find $Z_{\hat{B},I}>0$ in sub-filaments with $N({\rm{}H_2})<10^{23}~\rm{}cm^{-2}$, and we obtain $Z_{\hat{B},I}<0$ in the Main Ridge with $N({\rm{}H_2})>10^{22}~\rm{}cm^{-2}$. A transition threshold at $N({\rm{}H_2})\sim{}10^{22}~\rm{}cm^{-2}$ is thus consistent with our data. A much higher threshold is inconsistent with $Z_{\hat{B},I}/\sigma(Z_{\hat{B},I})<-3$ on cloud scales at $N({\rm{}H_2})\ge{}3\times{}10^{22}~\rm{}cm^{-2}$ (Fig.~\ref{fig-app:IvsB-all}). Conversely, a much lower threshold is inconsistent with the cloud-scale trend of increasing $Z_{\hat{B},I}$ with decreasing $N({\rm{}H_2})$ and the fact that $Z_{\hat{B},I}/\sigma(Z_{\hat{B},I})>3$ at $N({\rm{}H_2})=10^{22}~\rm{}cm^{-2}$ in the sub-filaments dominating the low-density envelope of DR21 (Fig.~\ref{fig:relative-orientations_BGI}~[left]). 

The fact that the alignment between $\hat{B}_{\rm pos}$ and $\hat{\nabla}_{\perp}I$ at a given column density depends on the environment (i.e., the $\langle{}z_{i,\hat{B},I}\rangle$ differ between the Main Ridge and sub-filaments) demonstrates that column density alone cannot fully characterize the relationship between the magnetic field and cloud density structure. Additional physical properties must play a role. The scatter in the $z_{i,\hat{B},I}$ observed for a given region and column density range points in the same direction: column density is not the core factor determining the relative orientation between the magnetic field and the cloud density structure. There are substantial pixel-to-pixel variations at fixed column density within a 
given region, and the \emph{ensemble} properties of these pixels as, e.g.,measured by $Z$ and $\langle{}z_i\rangle$ are the key observables for subsequent analysis. The $z_i$ in 
Fig.~\ref{fig:relative-orientations_BGI} tend to cluster at the extreme ends of their value range, near $|z_i|\approx{}1$. We caution that this is at least in part a consequence of how the $z_i$ are defined: the ordinate $z_i = \cos(2\cdot\Phi_i)$ has a non-linear connection to the angles $\Phi_i$, so that even a uniformly random sample of $\Phi_i$ 
would produce clustering near $|z_i|\approx{}1$. Our analysis relies on the averages $\langle{}z_i\rangle$, and a deeper interpretation of the $z_i$ distribution is beyond the scope of this paper.

\subsection{Relative Alignment between Cloud Structure and Gravitational Acceleration\label{sec:G-vs-density}}
In Figure~\ref{fig:relative-orientations_BGI}~(middle), we explore the extent to which the direction of gravitational acceleration, $\vec{g}_{\rm{}pos}$, drives the orientation of intensity and density gradients in DR21. The relative orientation between $\vec{g}_{\rm{}pos}$ and $\hat{\nabla}_{\perp}I$ is characterized through the variables $Z_{\hat{g},I}$ and $z_{i,\hat{g},I}$.

Trends between $\vec{g}_{\rm{}pos}$ and $\hat{\nabla}_{\perp}I$ closely mirror those between $\hat{B}_{\rm{}pos}$ and $\hat{\nabla}_{\perp}I$ discussed in Sec.~\ref{sec:B-vs-density}. Values $Z_{\hat{g},I}/\sigma_{\rm{}gen}(Z_{\hat{g},I})>3$ in sub-filaments indicate that $\vec{g}_{\rm{}pos}$ is aligned with elongated cloud structure in these regions, while the Main Ridge is characterized by $Z_{\hat{g},I}/\sigma_{\rm{}gen}(Z_{\hat{g},I})<-3$, implying that local filamentary cloud segments are oriented perpendicular to $\vec{g}_{\rm{}pos}$. As in the case of $\hat{B}_{\rm{}pos}$ vs.\ $\hat{\nabla}_{\perp}I$, the values of $Z_{\hat{g},I}$ and $\langle{}z_{i,\hat{g},I}\rangle$ differ between sub-filaments and the Main Ridge by much more than the respective uncertainties $\sigma_{\rm{}gen}(Z_{\hat{g},I})$ and $\sigma_{\rm{}gen}(\langle{}z_{i,\hat{g},I}\rangle)$. Substantial scatter in the $z_{i,\hat{g},I}$ for a given 
region and column density range again indicates pixel-to-pixel variations in this property, as discussed for $z_{i,\hat{B},I}$ in Sec.~\ref{sec:B-vs-density}.
The environment-dependent alignment between $\vec{g}_{\rm{}pos}$ and $\hat{\nabla}_{\perp}I$ implies that the orientation of $\hat{\nabla}_{\perp}I$ alone cannot predict the direction of $\vec{g}_{\rm pos}$. Some additional process must govern the overall structure of clouds, beyond what is captured by the gravitational field alone.

\subsection{Relative Alignment between Magnetic Field and Gravitational Acceleration\label{sec:BvsG}}
In Figure~\ref{fig:relative-orientations_BGI}~(right), we explore the correlation between the orientations of the projected magnetic field and the gravitational acceleration, as characterized through the variables $Z_{\hat{B},\hat{g}}$ and $z_{i,\hat{B},\hat{g}}$. The trend in the orientations $\hat{B}_{\rm{}pos}$ vs.\ $\vec{g}_{\rm{}pos}$ fundamentally differs from the trends in $\hat{B}_{\rm{}pos}$ vs.\ $\hat{\nabla}_{\perp}I$ and $\vec{g}_{\rm{}pos}$ vs.\ $\hat{\nabla}_{\perp}I$ discussed in Secs.~\ref{sec:B-vs-density}--\ref{sec:G-vs-density}. Here we observe $Z_{\hat{B},\hat{g}}/\sigma_{\rm{}gen}(Z_{\hat{B},\hat{g}})>3$ in most of the cloud. This trend is independent of the cloud environment, i.e., $\hat{B}_{\rm{}pos}$ and $\vec{g}_{\rm{}pos}$ are aligned both in the sub-filaments and the Main Ridge.

The orientation of $\vec{g}_{\rm{}pos}$ is thus a good predictor of the orientation of $\hat{B}_{\rm{}pos}$. This suggests that $\vec{B}$ and $\vec{g}$ are connected, either directly, in the sense that one vector field shapes the other, or indirectly, in the sense that both fields are shaped by a common underlying process. The strong, simple, and environment-independent character of this alignment, in contrast to the environment-dependent trends seen for $\hat{B}_{\rm{}pos}$ vs.\ $\hat{\nabla}_{\perp}I$ and $\vec{g}_{\rm{}pos}$ vs.\ $\hat{\nabla}_{\perp}I$, suggests that the connection between $\hat{B}_{\rm{}pos}$ and $\vec{g}_{\rm{}pos}$ reflects a fundamental physical process in the formation and evolution of molecular cloud structure.

Scatter in the $z_{i,\hat{B},\hat{g}}$ for a given region and column density range again indicates substantial pixel-to-pixel variations in the relative orientation of $\hat{B}_{\rm pos}$ vs.\ $\vec{g}_{\rm pos}$. However, the $z_{i,\hat{B},\hat{g}}$ cluster near $+1$ independently of the region, indicating that the connection between $\hat{B}_{\rm pos}$ and $\vec{g}_{\rm pos}$ is closer than those between the orientation pairs studied above.

\section{Interpretation and Discussion}\label{sec:interpretation}

\subsection{DR21: An Ordered Magnetic Field with Local Complexity}
The magnetic field structure of DR21, as characterized in Sec.~\ref{sec:data-analysis} and Fig.~\ref{fig:relative-orientations_BGI},  can be summarized as follows. First, the average properties of the magnetic field change smoothly with column density (e.g., $Z_{\hat{B},I}$ vs.\ column density in a given region) and systematically between cloud regions (e.g., the difference in $Z_{\hat{B},I}$ between sub-filaments and Main~Ridge at fixed column density). Second, there are substantial pixel-to-pixel variations within all of these regions, alongside 
region-to-region variations. Examples include the general scatter in the $z_i$ within any region at fixed column density (Fig.~\ref{fig:relative-orientations_BGI}), and the region-to-region differences between $Z$ and $\langle{}z_i\rangle$ at the same column density.

The data in particular indicate that \emph{the column density is not closely connected to changes in magnetic field structure}. We stress this because numerous studies, including the present one, explore how the relative orientation between $\hat{B}_{\rm{}pos}$ and 
$\hat{\nabla}_{\perp}I$ changes as a function of column density (following \citealt{Soler2013:imprint-magnetic-field}, \citealt{Planck2016:XXXV_HRO}, \citealt{Soler2019:HerschelPlanckMagneticField}). We agree with previous work that the relative alignment between $\hat{B}_{\rm{}pos}$ and 
$\hat{\nabla}_{\perp}I$ shifts from preferentially parallel at low column density to preferentially perpendicular at high column density. However, there is region-to-region dependence in the column density at which this transition occurs (seen in $Z_{\hat{B},I}$ and $\langle{}z_{i,\hat{B},I}\rangle$), and there is pixel-to-pixel variation in the alignment at fixed column density within a given region (seen in the scatter among the $z_{i,\hat{B},I}$; see Sec.~\ref{sec:B-vs-density} for details).

There clearly is value in using column density as a parameter encoding changes in magnetic field structure. However, substantial and important detail is lost when treating it as the sole parameter. Theoretical investigations should acknowledge pixel-to-pixel and region-to-region variations at fixed column density when modeling magnetic field structure in molecular clouds.

The nature of pixel-to-pixel and region-to-region variations in magnetic field structure at fixed column density is beyond the scope of our study. We speculate that small-scale perturbations from feedback, regime transitions between cloud-scale and localized clump-scale dynamics (e.g., in turbulence and gravitation), the stochastic behavior of MHD waves, and projection of the three-dimensional magnetic field onto two on-sky dimensions are among the factors contributing to this scatter.

\subsection{Importance of Magnetic Fields for Present-Day Stability against Collapse}\label{sec:stability-turbulence}
Understanding the role of magnetic fields in the DR21 
region requires considering the balance between 
self-gravity, magnetic support, and turbulent pressure. 
Here we summarize the key physical arguments; a more 
detailed treatment is provided in Appendix~\ref{sec-app:stability-turbulence}.

The DR21 Main Ridge is unstable to gravitational collapse unless it is supported by strong magnetic fields. This follows from the fact that the observed mass-to-length ratio of $M/\ell=3,650\,M_{\odot}\,{\rm{}pc^{-1}}$ is observed (Sec.~\ref{sec:segmentation}) substantially exceeds the critical value of $(M/\ell)_{\rm{}cr,kin}\le{}817\,M_{\odot}\,{\rm{}pc^{-1}}$ (Appendix~\ref{sec-app:stability-turbulence}) holding for infinite cylinders in hydrostatic equilibrium that are supported by isothermal pressure and random ``turbulent'' gas motions \citep{ostriker1964:polytropes}. The situation is less clear for sub-filaments, as the observed mass-to-length ratios of $164\,M_{\odot}\,{\rm{}pc^{-1}}$ to $556\,M_{\odot}\,{\rm{}pc^{-1}}$ (Table~\ref{tab:subfilaments}) do not differ significantly from critical values between $\le{}150\,M_{\odot}\,{\rm{}pc^{-1}}$ and $\le{}500\,M_{\odot}\,{\rm{}pc^{-1}}$ that strongly depend on the uncertain gas velocity dispersion in these cloud features.

\begin{table*}
\caption{Stability Analysis for the DR21 Main Ridge (Sec.~\ref{sec:stability-turbulence}).\label{tab:stability-analysis}}
\centering
\begin{tabular}{llllll}
\hline\hline
$N_{\rm{}H_2,lim}$ & $w$ &
    $\langle{}N_{\rm{}H_2}\rangle$ & $\langle{}n_{\rm{}H_2}\rangle$ & 
    $\langle{}B\rangle$ &
    \multirow{2}{*}[0.5ex]{$\displaystyle{}\frac{M/\Phi_B}{(M/\Phi_B)_{\rm{}cr}}$}\\
($10^{22}~\rm{}cm^{-2}$) & (pc) & ($10^{23}~\rm{}cm^{-2}$) &
    ($10^4~\rm{}cm^{-3}$) & ($\rm{}\mu{}G$) \\
\hline
2.5 & 1.2 & 1.0 & 2.7 & 285 & 1.7\\
10 & 0.5 & 1.5 & 9.4 & 643 & 1.1\\
\hline

\end{tabular}
\tablecomments{Widths $w$ and the mean column densities $\langle{}N_{\rm{}H_2}\rangle$ are determined for regions within column density boundaries $N_{\rm{}H_2,lim}$. These measurements allow to estimate mean densities $\langle{}n_{\rm{}H_2}\rangle$, mean magnetic field strengths $\langle{}B\rangle$, and the resulting mass-to-flux ratio $(M/\Phi_B)/(M/\Phi_B)_{\rm{}cr}$. See Appendix~\ref{sec-app:stability-turbulence} for details.}
\end{table*}

Scaling relations thought to generally describe the magnetized interstellar medium suggest that the DR21 Main Ridge is likely in a slightly magnetically supercritical state. To show this, we estimate the mean gas volume densities $\langle{}n_{\rm{}H_2}\rangle$ in the Main Ridge reported in Table~\ref{tab:stability-analysis}, where $\langle{}n_{\rm{}H_2}\rangle$ is estimated by dividing the observed mean column density $\langle{}N_{\rm{}H_2}\rangle$ by the observed width $w$ of the Main Ridge (see Appendix~\ref{sec-app:stability-turbulence}). We do this for two different region selections in the Main Ridge, implemented by considering cloud area above a column density $N_{\rm{}H_2,lim}$. The reference magnetic field strength for the dense interstellar medium,
\begin{equation}
B =
    B_0 \cdot
    \left(
        n_{\rm{}H_2} / 10^4~{\rm{}cm^{-3}}
    \right)^{\alpha}
\label{eq:Crutcher-relation}
\end{equation}
with $B_0\le{}150~\rm{}\mu{}G$ and $\alpha=0.65$ \citep{crutcher2012:review}, gives magnetic flux densities of 250 to $650~\rm{}\mu{}G$ for these gas volume densities. These expected values are consistent with CN Zeeman observations, which trace dense gas with effective excitation densities of a few $\times 10^4$\,cm$^{-3}$ for $T_{\rm gas} \approx 10$--$20$\,K \citep{Shirley2015}, towards DR21(OH) \citep{Falgarone2008}. Combined with the mean column densities $\langle{}N_{\rm{}H_2}\rangle$ given in Table~\ref{tab:stability-analysis}, the expected and critical mass-to-flux ratios may be compared through
\begin{equation}
\frac{M/\Phi_B}{(M/\Phi_B)_{\rm{}cr}} = 
    0.51 \cdot
    \left(
        \frac{
            \langle{}N_{\rm{}H_2}\rangle
            }{
            10^{23}~\rm{}cm^{-2}
            }
    \right) \cdot
    \left(
        \frac{\langle{}B\rangle}{1~\rm{}mG}
    \right)^{-1}
\label{eq:mass-to-flux_cr}
\end{equation}
(\citealt{Tomisaka2014:filament-lateral-field}; this critical mass-to-flux ratio for filaments is by a factor 1.5 larger than the value for initially spherical clouds from \citealt{Nakano1978:mass-to-flux}). This yields 
$(M/\Phi_B)/(M/\Phi_B)_{\rm{}cr}=1~\text{to}~2$, indicating a critical or mildly supercritical state (see Table~\ref{tab:stability-analysis} for all quantitative results listed above). This is broadly similar to the ratios $(M/\Phi_B)/(M/\Phi_B)_{\rm{}cr}=1.6~\text{to}~2.6$ found by \citeauthor{Ching2022:BISTRO-DR21}~(\citeyear{Ching2022:BISTRO-DR21}; we quote values after scaling their parameter $\lambda$ to match Eq.~[\ref{eq:mass-to-flux_cr}]), who estimate $\langle{}B\rangle$ through the Davis-Chandrasekhar-Fermi method, i.e., using the beam-to-beam variation of $\hat{B}_{\rm{}pos}$. A supercritical state is also indicated by the presence of star formation in DR21.

All these estimates have substantial systematic uncertainties, as the cloud density structure and other relevant parameters are only calculated in an approximate sense. Still, these estimates are consistent with the DR21 Main Ridge being in a moderately supercritical state, $(M/\Phi_B)/(M/\Phi_B)_{\rm{}cr}>1$, but \emph{not} in a drastically supercritical state, $(M/\Phi_B)/(M/\Phi_B)_{\rm{}cr}\gg{}1$.

\subsection{Birth of DR21 out of strongly-magnetized diffuse Precursor Gas\label{sec:strong-initial-B}}

In Sec.~\ref{sec:B-vs-density}, we show that elongated 
low-density structures in DR21 are preferentially aligned with 
$\hat{B}_{\rm pos}$, while elongated higher-density structures 
are preferentially perpendicular to it 
(Fig.~\ref{fig:relative-orientations_BGI}, left). Such 
transitions in relative orientation between density structures and 
magnetic fields have been studied systematically since 
\citet{Soler2013:imprint-magnetic-field} (see 
\citealt{Pattle2023:PPVII_MagneticFieldsCloudsCores} for a recent 
review.) The emerging consensus is that a transition from parallel 
at low density to perpendicular at high density requires that 
turbulent motions in the diffuse precursor gas did not greatly 
exceed the magnetic energy --- i.e., 
$\mathcal{M}_{\rm A} \lesssim 1$ at early times. This is not a 
sharp threshold: \citet{Soler2013:imprint-magnetic-field}, for 
example, recover such a transition even at an initial 
$\mathcal{M}_{\rm A} = 3$. However, the transition disappears 
for $\mathcal{M}_{\rm A} \gg 1$, though the precise boundary 
remains difficult to establish
(Fig.~10 of \citeauthor{Pattle2023:PPVII_MagneticFieldsCloudsCores}).

The low initial Alfv\'en Mach number implies the existence of a moderately-ordered background magnetic field in the initial state, for example characterized by a small position angle dispersion $\sigma(\Phi_i)$ at that time. This follows from the relation $\sigma(\Phi_i)\approx{}\mathcal{M}_{\rm{}A}$ prevailing in magnetized gas (\citealt{ostriker2001:turbulent-clouds}, their Eq.~[16]). Assuming an initially homogeneous gas density distribution subject to such a structured magnetic field, \citet{Mestel1985:PP_II} demonstrates that gaseous structures of a size below some critical scale will not collapse under self-gravity. This is also captured by Eq.~(\ref{eq:mass-to-flux_cr}), which states for such conditions that cloud structure below a certain column density threshold is magnetically subcritical, $M/\Phi_B<(M/\Phi_B)_{\rm{}cr}$. Provided the DR21 Main Ridge condensed out of the diffuse interstellar medium and increased its column density over time, this implies that the Main Ridge was also born as a subcritical structure. This picture is also consistent with simulations by \citet{Seifried2020:parallel-perpendicular}. That team suggests that $\vec{B}$ becomes perpendicular to cloud structure only in magnetically subcritical gas, $M/\Phi_B\lesssim{}(M/\Phi_B)_{\rm{}cr}$.

The present-day moderately supercritical state estimated in Sec.~\ref{sec:stability-turbulence}, combined with the initially subcritical state inferred here, implies that the DR21 Main Ridge must have accumulated mass relative to its magnetic flux over the course of its evolution. The following sections examine the mechanism by which this transition may have occurred.

\subsection{Efficient Accretion along Magnetic Field Lines}\label{sec:guided-accretion_rates}

Accretion flows aligned with the magnetic field are expected to proceed efficiently, because the Lorentz force, acting perpendicular to $\vec{B}$, cannot decelerate motions along field lines. Such flows will preferentially follow the gravitational acceleration $\vec{g}$. In DR21 we find that $\vec{g}_{\rm pos}$ is aligned with $\hat{B}_{\rm pos}$ in the sub-filaments (Sec.~\ref{sec:BvsG}, Fig.~\ref{fig:relative-orientations_BGI}~[right]), consistent with a geometry in which magnetic fields cannot impede gas flows. Section~\ref{sec:guided-accretion_retardation} demonstrates that the free-fall accretion rates expected in this scenario are quantitatively consistent with those necessary to fuel star formation and the formation of the Main Ridge. We also find that the sub-filaments themselves are aligned with $\hat{B}_{\rm pos}$ (Sec.~\ref{sec:B-vs-density}, Fig.~\ref{fig:relative-orientations_BGI}~[left]), a configuration seen in MHD simulations \citep{Nakamura2008:Taurus} and consistent with interpreting the sub-filaments as close  accretion streams.

The inferred alignment between $\vec{B}$ and $\vec{g}$ is the key observational result of this paper. It motivates the discussion that follows and drives our central astrophysical conclusions.

A key consequence of magnetically guided accretion is that it
increases the mass-to-flux ratio $M/\Phi_B$ at the locations
where material accumulates: motion along $\vec{B}$ adds mass to
a given volume without changing the magnetic flux threading it.
We quantify this effect in Table~\ref{tab:stability-analysis}.
Taking the sub-filament masses $M$ and widths $w$ from
Table~\ref{tab:subfilaments} (assuming $w \approx 0.5$~pc,
twice the \emph{Herschel} FWHM beam, since the sub-filaments
are marginally resolved), we estimate the mass surface density
along the sub-filament axes as $\Sigma = M/w^2$. Converting to
column density via $\Sigma = 2.8\,m_{\rm p}\,N_{\rm H_2} =
2{,}233\,M_{\odot}\,\mathrm{pc}^{-2}\,(N_{\rm H_2}/10^{23}\,
\mathrm{cm}^{-2})$, where a mean molecular weight per
$\mathrm{H}_2$ molecule of $\mu_{\mathrm{H}_2} =
2.8\,m_{\rm p}$ is assumed
\citep{kauffmann2008:mambo-spitzer}, yields column densities of
order $10^{23}\,\mathrm{cm}^{-2}$ and volume densities of order
$10^4\,\mathrm{cm}^{-3}$
(Table~\ref{tab:subfilaments}). These values are comparable to
the mean column density of the DR21 Main Ridge
(Table~\ref{tab:stability-analysis}), implying that accretion
of a single sub-filament would roughly double the enclosed mass
$M$ while leaving $\Phi_B$ unchanged. \emph{Accretion of a
single sub-filament along a magnetic field line would therefore
locally increase $M/\Phi_B$ by a factor $\sim 2$, drastically
reducing the ability of the magnetic field to support the
structure against collapse.}

\begin{figure*}
\centering
\includegraphics[width=\linewidth]{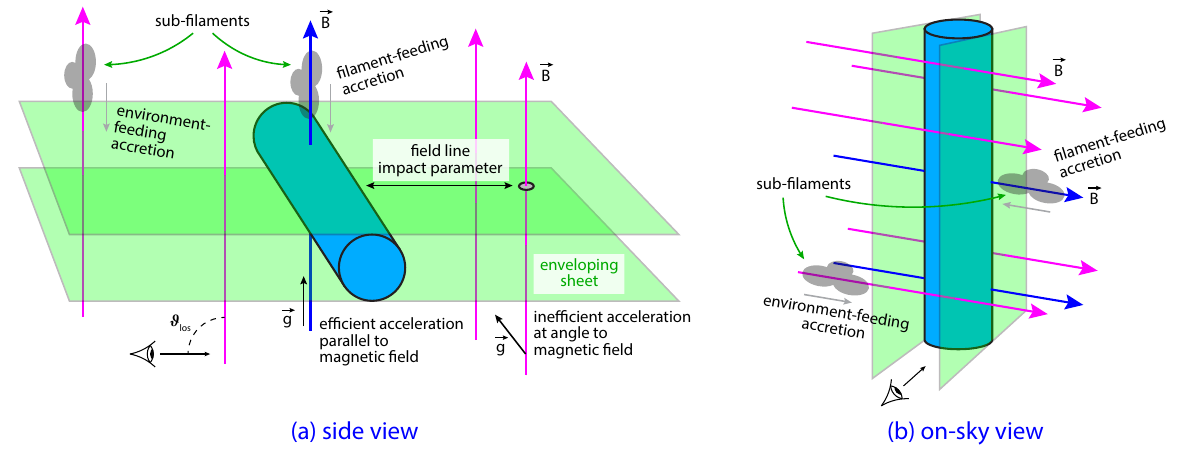}
\caption{Outline of a magnetic field and cloud geometry
consistent with the HAWC+ observations. \emph{Blue and magenta vertical
arrows} indicate the global structure of the magnetic field.
The \emph{blue cylinder} represents a dense cloud, such as the
DR21 Main Ridge. The \emph{green layers} represent a sheet that
envelopes the dense cloud and is oriented perpendicular to the
magnetic field. \emph{Panel~(a)} presents a view in which the
\emph{observer} is located towards the left of the figure; the
line of sight in this model is nearly perpendicular to
$\vec{B}$. \emph{Gray cloudlets} represent gas parcels
following trajectories parallel to $\vec{B}$ that accrete onto
the sheet or the dense cloud, depending on the impact parameter
of the field line they follow. \emph{Acceleration vectors} span
non-zero angles with the local field line for non-zero impact
parameters. \emph{Panel~(b)} provides a rotated view closely
aligned with the observer's
perspective.\label{fig:peripheral-accretion}}
\end{figure*}

\subsection{Observational Kinematics of Magnetized
Accretion}\label{sec:guided-accretion_retardation}

The accretion model developed in
Sec.~\ref{sec:guided-accretion_rates} makes quantitative
predictions for both mass infall rates and velocities that can
be tested versus observed gas kinematics. We begin with the
rates, then turn to velocities.

\paragraph{Rates} In Appendix~\ref{sec-app:accretion} we estimate that DR21 forms
stars at a rate $\gg 10^{-3}\,M_{\odot}\,\mathrm{yr}^{-1}$,
based on star formation rates typical for dense molecular clouds
\citep{lada2010:sf-efficiency}, and that accretion rates
$\sim{}5\times{}10^{-3}\,M_{\odot}\,\mathrm{yr}^{-1}$ were
necessary to form the DR21 Main Ridge during the cloud formation
time indicated by stellar ages \citep{Beerer2010:Cygnus-X}.
Assuming that the sub-filaments reside within $\sim{}2~\rm{}pc$
of the Main Ridge (Fig.~\ref{fig:main}; given projected lengths
$\sim{}2~\rm{}pc$ in Table~\ref{tab:subfilaments}), they will
accrete into the Main Ridge within
$t_{\rm{}accr}\sim{}4\times{}10^5~\rm{}yr$
(Appendix~\ref{sec-app:accretion}). Given sub-filament masses
$M\approx{}500\,M_{\odot}$ (Table~\ref{tab:subfilaments}), such
motions would result in accretion rates
$M/t_{\rm{}accr}\approx{}1.3\times{}10^{-3}\,M_{\odot}\,
{\rm{}yr^{-1}}$ per sub-filament. With several sub-filaments
present, the combined accretion rate is sufficient to sustain
both star formation and the continued growth of the Main Ridge.
The presently observed sub-filaments represent a snapshot of
this ongoing process; their combined mass
($\sim{}3{,}800\,M_{\odot}$, Table~\ref{tab:subfilaments})
accounts for $\sim{}20\%$ of the Main Ridge mass, consistent
with a picture in which successive generations of sub-filaments
fed the Main Ridge over its lifetime of several
$10^6~\rm{}yr$. We further argue in
Appendix~\ref{sec-app:evolution-subfilaments} that the
sub-filaments are unlikely to collapse under their own
self-gravity before merging with the Main Ridge, implying that
star formation in the DR21 region is confined mostly to the Main Ridge.

\paragraph{Infall velocities} Free-fall towards filamentary
molecular clouds has been considered by
\citet{heitsch2013:infall-onto-filaments}, who show that infall
motions follow
\begin{equation}
v(r) = -2 \left[ G \, \frac{M}{\ell} \,
    \ln\!\left(\frac{r_{\rm ini}}{r}\right)
    \right]^{1/2} ,
\label{eq:model-Heitsch}
\end{equation}
where $M/\ell$ is the mass-per-unit-length of the cloud and
motions are assumed to start from rest at radius $r_{\rm ini}$.
A quantitative evaluation is provided in
Appendix~\ref{sec-app:accretion}, yielding characteristic
accretion velocities $\approx 8\,\mathrm{km\,s}^{-1}$ for the
DR21 Main Ridge.

Such fast motions are, however, inconsistent with radial gas
velocities of $\approx (2\pm 1)\,\mathrm{km\,s}^{-1}$ relative
to the systemic velocity of the Main Ridge observed by
\citet{schneider2010:dr21} and \citet{Bonne2023:DR21}. These
authors study radial gas motions in the vicinity of the Main
Ridge, e.g., through position--velocity diagrams, and do not
detect velocity gradients consistent with the
$\approx 8\,\mathrm{km\,s}^{-1}$ acceleration expected for
free fall.

We therefore refine our picture of magnetized accretion flows in
DR21 as illustrated in Fig.~\ref{fig:peripheral-accretion}. The
figure presents a model of the environment within which DR21
resides, inspired by the MHD simulations of
\citet{Nakamura2008:Taurus}. Assuming a strong, ordered
large-scale magnetic field, gas accretes into a sheet oriented
perpendicular to $\vec{B}$. A dense structure like the DR21 Main
Ridge forms within this sheet and is therefore oriented
perpendicular to the field. Lower-density structures accreting
into the sheet are stretched along the field direction; within
the model, the observed sub-filaments correspond to such
accretion streams.

A key parameter is the angle $\vartheta_{\rm los}$ between the
background magnetic field and the line of sight.
Figure~\ref{fig:peripheral-accretion} is drawn for the case
$\vartheta_{\rm los} \approx 90\degr$, as will become relevant
below. Panel~(b) shows a view close to the observer's
perspective: the Main Ridge (blue cylinder) is permeated by a
magnetic field oriented roughly East--West, and the
sub-filaments are arranged along the field lines. Panel~(a)
provides a rotated view that illustrates the three-dimensional
configuration along the line of sight.

Because accretion motions are confined to directions parallel to
$\vec{B}$, only a fraction of the true velocity manifests as a
line-of-sight component: $|v_{\rm los}| = |\cos(\vartheta_{\rm
los})|\,|\vec{v}|$. For DR21, an angle $\vartheta_{\rm los} =
75\arcdeg$ would reduce the free-fall speed of $|\vec{v}| =
8\,\mathrm{km\,s}^{-1}$ to $|v_{\rm los}| =
2\,\mathrm{km\,s}^{-1}$, consistent with the kinematics
observed by \citet{schneider2010:dr21}.

\citet{Bonne2023:DR21} report an important subtlety that
suggests a refinement of the model shown in
Fig.~\ref{fig:peripheral-accretion}. Their [\cii] data reveal
that gas on both the eastern and western sides of the Main Ridge
is systematically redshifted relative to the systemic velocity
of the Ridge. In the simplest version of our model
(Fig.~\ref{fig:peripheral-accretion}~[a]), a parcel moving
parallel to $\vec{B}$ from below the enveloping sheet has
radial velocity $v_{\rm rad,+} = v\,\sin(\vartheta_{\rm pos})$,
while a parcel moving antiparallel to $\vec{B}$ from above has
$v_{\rm rad,-} = -v\,\sin(\vartheta_{\rm pos})$, where
$\vartheta_{\rm pos} = \vartheta_{\rm los} - 90\arcdeg$. The
systematic redshift on \emph{both} sides of the Main Ridge is
inconsistent with this prediction, as it requires
$v_{\rm rad,-} \neq -v_{\rm rad,+}$. As
\citet{Bonne2023:DR21} demonstrate (following
\citeauthor{Bonne2020:Musca}~\citeyear{Bonne2020:Musca},
their Fig.~22), this observation can be explained by
magnetically guided accretion along \emph{curved} field lines,
so that $\vartheta_{\rm los}$ (and thus $\vartheta_{\rm pos}$)
varies across the observed region. \citet{Bonne2023:DR21} argue
that this curvature encodes important information about the
formation history of the Main Ridge, building on a model by
\citet{Inoue2018}.

Including limited magnetic field curvature does not affect our
main conclusion: the model of magnetically guided accretion onto
the DR21 Main Ridge requires the magnetic field to lie roughly
in the plane of the sky ($\vartheta_{\rm los} \approx 90\degr$).
Values much closer to zero would not suffice to explain the
observed infall kinematics as a projection effect.

The discussion so far has focused on accretion flows directed
into the Main Ridge. However, the model of
Fig.~\ref{fig:peripheral-accretion} also accommodates gas
reaching the sheet at locations far from the central structure,
along field lines with large impact parameters---a geometry
consistent with trends seen in the simulations of
\citet{Nakamura2008:Taurus}. For such flows, the
three-dimensional separation from the Main Ridge can be much
larger than the projected on-sky distance, substantially
reducing the depth of the gravitational potential those flows
experience. The resulting accretion velocities are therefore
well below those predicted by
Eq.~(\ref{eq:model-Heitsch}), offering a further explanation
for the modest observed infall speeds.

Interestingly, \citet{Dobashi2019:DR21} and \citet{Schneider2023} propose the presence of several distinct
clouds along the line of sight towards DR21, which could in
principle constitute additional structures within the sheet
depicted in Fig.~\ref{fig:peripheral-accretion}. These authors
base their suggestion on spatially overlapping cloud components
seen at radial velocities $\gtrsim 10\,\mathrm{km\,s}^{-1}$
above the systemic velocity of the Main Ridge. Such
high-velocity components are likely important for understanding
the DR21 complex as a whole, but they fall outside the scope of
the model in Fig.~\ref{fig:peripheral-accretion}, which is
designed to explain the kinematics of the Main Ridge and its
immediately surrounding sub-filaments. Structures accreting
along field lines with large impact parameters would arrive at
velocities close to that of the Main Ridge, not at offsets of
$\gtrsim 10\,\mathrm{km\,s}^{-1}$, and therefore cannot
account for the high-velocity components identified by
\citeauthor{Dobashi2019:DR21} and \citeauthor{Schneider2023}.

\subsection{Collapse of a magnetized Structure in Sub-Filament~S}\label{sec:bending-magnetic-fields}
In sub-filament~S (see Figure~\ref{fig:masks}), the magnetic field transitions from
predominantly perpendicular to the sub-filament axis south of
20:39:10 $+42$:14:00 (J2000) to increasingly parallel north
of this location (Sec.~\ref{sec:structure-overview}), consistent
with substantial dragging of an initially perpendicular field
by an accretion flow as proposed by \citet{Pillai2020}. The underlying magnetic
flux density can be gauged within this model.

We base our estimate on the steady-state advection-diffusion model of \citet{Tapinassi2024:field-dragging}, developed to explain the field morphology observed by \citet{Pillai2020}.  They consider gas at density $n_{\rm{}H_2}$ flowing through a filament of width $w$ that is initially permeated by a homogeneous magnetic field that is oriented perpendicular to the filament. This magnetic field then bends as gas flows along the filament at velocity $v_{\rm{}flow}$. 
Assuming the presence of ambipolar diffusion,
\citeauthor{Tapinassi2024:field-dragging} show that, in
steady state, the magnetic field morphology is governed by
a single dimensionless parameter, the ambipolar diffusion
Reynolds number,
\begin{equation}
\begin{split}
\mathcal{R}_{\rm{}ad}
    &= \frac{1}{2}
        \cdot \left( \frac{\chi}{2} \right)
        \cdot \left( \frac{w}{0.1~\rm{}pc} \right)
        \cdot \left( \frac{v_{\rm{}flow}}{\rm{}km\,s^{-1}} \right) \\
    &\quad\quad\quad
        \cdot \left( \frac{n_{\rm{}H_2}}{10^4~\rm{}cm^{-3}} \right)^{3/2}
        \cdot \left( \frac{B_{\rm{}c}}{100~\rm\mu{}G} \right)^{-2} \, ,
\end{split}
\label{eq:R_ad}
\end{equation}
with $\chi=1~\text{to}~3$ parameterizes the uncertainty in the
ion-neutral coupling \citep[e.g.,][]{Pinto2008, Padovani2009}, and $B_{\rm{}c}$ being the magnetic flux density on the filament's central axis. This can be rearranged to estimate the magnetic flux density, which depends on the other properties as $B_{\rm{}c}\propto{}(\chi\cdot{}w\cdot{}v_{\rm{}flow})^{1/2}\cdot{}\mathcal{R}_{\rm{}ad}^{-1/2}\cdot{}n_{\rm{}H_2}^{3/4}$. Alternatively, flow speeds estimated from $\mathcal{R}_{\rm{}ad}$ scale as $v_{\rm{}flow}\propto{}\mathcal{R}_{\rm{}ad}\cdot{}\chi^{-1}\cdot{}w^{-1}\cdot{}n_{\rm{}H_2}^{-3/2}\cdot{}B_{\rm{}c}^2$.

Following Fig.~2 of \citet{Tapinassi2024:field-dragging}, flows can cause noticeable magnetic field deflections by angles $\ge{}20\arcdeg$ if $\mathcal{R}_{\rm{}ad}\gtrsim{}1/2$, and we thus adopt this as a lower limit on $\mathcal{R}_{\rm{}ad}$. \citet{Hu2021:DR21-accretion} study sub-filament~S in more detail and find $v_{\rm{}flow}=3.7~\rm{}km\,s^{-1}$ and $n_{\rm{}H_2}=1.6\times{}10^4~\rm{}cm^{-3}$, while their Fig.~2 suggests $w\approx{}0.3~\rm{}pc$. Substituting these values in Eq.~(\ref{eq:R_ad}) yields $B_{\rm{}c}\lesssim{}475~\rm{}\mu{}G$, with a systematic uncertainty by at least a factor of two due to uncertainties in $\chi$ and other physical properties. This is broadly consistent with the scaling relation Eq.~(\ref{eq:Crutcher-relation}), which gives $B\le{}200~\rm{}\mu{}G$ for this density.

\subsection{High-Mass Star Formation Magnetic Field}
\citet{Soler2013:imprint-magnetic-field} identify the column density at which the cloud density structure transitions from being parallel to $\hat{B}_{\rm{}pos}$ to being perpendicular to $\hat{B}_{\rm{}pos}$ as an important parameter. Our analysis in Sec.~\ref{sec:B-vs-density} shows that a threshold at $N({\rm{}H_2})\sim{}2\times{}10^{22}~\rm{}cm^{-2}$ in DR21 is consistent with our data. Interestingly, this is consistent with, but at the higher end of, the transition column densities in the range of $(0.4~{\rm{}to}~5)\times{}10^{22}~\rm{}cm^{-2}$ as \citet{Planck2016:XXXV_HRO} found for nearby molecular clouds using \emph{Planck} (excluding the Corona~Australis cloud, which is an outlier). These results need to be interpreted with some care, given the instrument's low angular resolution and the fact that re-analysis of \emph{Planck} data by \citet{Soler2019:HerschelPlanckMagneticField} delivers markedly lower threshold column densities in the range $(0.2~ {\rm{}to}~1.3)\times{}10^{22}~\rm{}cm^{-2}$ for well-characterized regions. Still, the broad similarity in threshold column densities is remarkable because these clouds are of much lower density and are primarily sites of low-mass star formation, compared to the dense and massive DR21 region forming high-mass stars. This finding of similar threshold column densities in regions with different physical parameters has no immediate accepted interpretation \citep{Pattle2023:PPVII_MagneticFieldsCloudsCores}. It broadly indicates that the magnetic field plays similar roles in the assembly of regions of high-mass and low-mass star formation. A caveat is that the distinction between Main-Ridge-like and sub-filament-like environments highlighted throughout this work is likely less pronounced in low-mass star-forming regions, where 
density contrasts and gravitational potentials between the central ridge and surrounding sub-filaments are smaller. Direct tests of whether the column-density threshold and the environment-dependent trends seen in DR21 generalize to low-mass clouds will require comparable analyses across the broader SIMPLIFI sample, which will be presented in subsequent papers.

\section{Summary and Conclusions}\label{sec:summary}
We present \sofia/\hawc\ 214~$\mu$m polarimetric observations of the DR21 region, tracing the plane-of-sky magnetic field from the dense Main Ridge into its network of sub-filaments at $\sim$0.1~pc resolution. We characterize the relative orientations of the magnetic field ($\hat{B}_{\rm pos}$), the gravitational acceleration ($\vec{g}_{\rm pos}$), and the cloud density structure ($\hat{\nabla}_{\perp}I$), and interpret these in the context of magnetized cloud evolution. Our principal conclusions are the following.

\begin{itemize}

\item The DR21 region is permeated by an ordered magnetic field with substantial local complexity (Sec.~\ref{sec:interpretation}). The average properties of the field change smoothly and systematically between cloud regions defined by column density and other characteristics, but there are also substantial pixel-to-pixel variations within any given region. This indicates that no single parameter --- column density included --- fully encodes how magnetic field structure changes within molecular clouds, and theoretical work 
should not treat any single parameter as such.

\item The DR21 Main Ridge, with $M/\ell = 3{,}650\,M_{\odot}\,{\rm pc^{-1}}$, exceeds the critical mass-to-length ratio for support by thermal pressure and turbulence alone (Sec.~\ref{sec:stability-turbulence}). Magnetic fields are therefore essential to understanding its present dynamical state.

\item Elongated cloud structure transitions from being preferentially parallel to $\hat{B}_{\rm pos}$ in low-density sub-filaments to preferentially perpendicular in the dense Main Ridge. This transition, occurring at $N({\rm H_2}) \sim 2\times 10^{22}$~cm$^{-2}$, indicates that the DR21 complex formed from strongly magnetized precursor gas with $\mathcal{M}_{\rm A} \lesssim 1$ and an initially sub-critical mass-to-flux ratio (Sec.~\ref{sec:strong-initial-B}).

\item The transition column density in DR21, a high-mass star-forming region, is consistent with thresholds found in nearby low-mass star-forming clouds by \emph{Planck} (Sec.~\ref{sec:strong-initial-B}). This suggests that magnetic fields play a similar role in structuring molecular clouds across a wide range of star-forming environments.

\item The orientation of $\vec{g}_{\rm pos}$ is strongly aligned with $\hat{B}_{\rm pos}$ across the entire DR21 complex, independent of column density and environment (Sec.~\ref{sec:BvsG}). This contrasts with the environment-dependent alignment of both $\hat{B}_{\rm pos}$ and $\vec{g}_{\rm pos}$ relative to $\hat{\nabla}_{\perp}I$, and constitutes the central observational result of this paper.

\item This $\vec{g}_{\rm pos}$--$\hat{B}_{\rm pos}$ alignment is consistent with magnetically guided accretion, in which sub-filaments channel gas along field lines into the Main Ridge. Free-fall accretion from sub-filaments would deliver mass at rates of several $10^{-3}\,M_{\odot}\,{\rm yr^{-1}}$, sufficient to assemble the Main Ridge within a few $10^6~\rm{}yr$ and to fuel ongoing star formation at rates $\gg{}10^{-3}\,M_{\odot}\,{\rm yr^{-1}}$ (Sec.~\ref{sec:guided-accretion_rates}).

\item Accretion of a single sub-filament along field lines would roughly double the local mass-to-flux ratio, driving the transition from the initially sub-critical state to the mildly supercritical state observed today (Sec.~\ref{sec:guided-accretion_rates}).

\item Observed radial velocities $\approx{}2~\rm{}km\,s^{-1}$ fall well below speeds $\approx{}8~\rm{}km\,s^{-1}$ expected for free-fall. This discrepancy is naturally explained if accretion is confined to directions along a magnetic field that lies nearly in the plane of the sky ($\vartheta_{\rm los} \approx 75\arcdeg$), so that only a small fraction of the true velocity projects onto the line of sight (Sec.~\ref{sec:guided-accretion_retardation}).

\item The bending of magnetic field lines in sub-filament~S constrains the local field strength to $B \lesssim 475\,\mu{\rm G}$, consistent with the scaling relation at the relevant density (Sec.~\ref{sec:bending-magnetic-fields}).

\item The sub-filaments are unlikely to have engaged in significant star formation before accreting into the Main Ridge, as their local free-fall time ($t_{\rm ff} \approx 0.3$~Myr) is comparable to the accretion timescale ($t_{\rm accr} \sim 0.42$~Myr; Appendix~\ref{sec-app:evolution-subfilaments}). This supports their interpretation as coherent mass-feeding channels rather than independent star-forming structures.

\end{itemize}

This study presents first results from the SIMPLIFI survey. Forthcoming papers will analyze polarization fraction trends, derive magnetic field strengths via advanced modeling, and integrate molecular line kinematics. Together with observations of the broader SIMPLIFI filament sample, these efforts will constrain the relative roles of turbulence, gravity, and magnetic fields in regulating star formation within filaments.

\begin{acknowledgments}
This paper is dedicated to the memory of Karl M. Menten, whose mentorship and friendship shaped the careers of the lead authors. We are grateful for everything he taught us. We thank Tao-Chung Ching for kindly providing the JCMT/POL-2 data from \citet{Ching2022:BISTRO-DR21}, and R.~Gutermuth and R.~Pokhrel for providing the \emph{Herschel}-based column density and temperature maps from \citet{pokhrel2020}. We thank the anonymous referee for a careful and constructive review that improved the clarity and presentation of this work. This paper is based on observations made with the NASA/DLR Stratospheric  Observatory for Infrared Astronomy (\sofia). \sofia\ was jointly operated by the Universities Space Research Association, Inc. (USRA), under NASA contract  NNA17BF53C, and the Deutsches SOFIA Institut (DSI) under DLR contract 50 OK 0901 to the University of Stuttgart.TGSP would like to thank all SOFIA observatory personnel for their support throughout the SIMPLIFI project, and in particular Peter Ashton, Simon Coud\'e, and Sarah Eftekharzadeh for the successful research 
flights leading to the results presented here. TGSP gratefully acknowledges support by NASA award \#09-0215 issued by USRA and awards from National Science Foundation under grant no.\ AST-2009842 and AST-2108989. This work was performed in part at the Jet Propulsion Laboratory, California Institute of Technology, under contract with the National Aeronautics and Space Administration (80NM0018D0004). 
DS acknowledges support of the Bonn-Cologne Graduate School, which is funded through the German Excellence Initiative as well as funding by the Deutsche Forschungsgemeinschaft (DFG) via the Collaborative Research Center SFB 1601 ‘Habitats of Massive Stars Across Cosmic Time’ (subprojects B1 and B4).
\end{acknowledgments}

\bibliography{simplifi_dr21}{}
\bibliographystyle{aasjournal}

\appendix

\section{Uncertainty of the Projected Rayleigh Statistic\label{app:Z-uncertainty}}
The PRS is expressed as the quantity $Z$:
\begin{equation}
Z = \frac{\sum_i^{n_{\rm{}ind}} \cos(2\cdot{}\Phi_i)}{(n_{\rm{}ind}/2)^{1/2}} \, .
\end{equation}
The significance of a result is given by $|Z|/\sigma(Z)$, where $\sigma(Z)$ is the standard deviation in $Z$ resulting from uncertainties. An extreme limit can be found in the case of random orientations between the respective pseudovectors, i.e., if the $\Phi_i$ are drawn randomly from the interval $[0\degr,90\degr]$ using a uniform probability distribution. \citet{jow2018} demonstrate that $\sigma(Z)=1$ in that case. This highlights an important aspect of the definition of $Z$, as this implies that $Z=Z/\sigma(Z)$. In other words, the value of the PRS is equal to its signal-to-noise ratio (SNR) under the assumption that the $\Phi_i$ are distributed randomly. 
For a fixed number of independent measurements $n_{\rm{}ind}$, the maximum possible value of $|Z|$ is
\begin{equation}
|Z|_{\rm{}max} = (2\cdot{}n_{\rm{}ind})^{1/2}
\end{equation}
since $|\sum_i^{n_{\rm{}ind}}\cos(2\cdot{}\Phi_i)|\le{}n_{\rm{}ind}$. This upper bound grows with $n_{\rm{}ind}$, reflecting the fact that adding independent measurements increases the attainable significance. However, real observations of $\Phi_i$ are rarely drawn from a uniform distribution, so that the actual SNR typically exceeds $|Z|$. \citet{jow2018} show that the uncertainty in $Z$ for angles drawn from a single parent distribution with a well-defined (and generally non-zero) mean is
\begin{equation}
\sigma_{\rm{}gen}(Z) =
    \left(
        \frac{
            2 \cdot \sum_i^{n_{\rm{}ind}} \cos^2(2 \cdot \Phi_i) - Z^2
        }{n_{\rm{}ind}}
    \right)^{1/2} \, ,
\label{eq-app:Z-unc-common_angle}
\end{equation}
which assumes that the $\Phi_i$ within a given region differ only because they are independent samples from this parent distribution, encompassing both observational scatter and intrinsic astrophysical variation. Because $\sigma_{\rm{}gen}(Z)$ captures the combined effect of measurement errors and real source structure, it is in some sense conservative, and we adopt it as our preferred measure of the uncertainty in $Z$.

We may alternatively estimate the extent to which measurements of $Z$ are influenced by pure observational errors in the angles $\Phi_i$, as characterized in the form of pixel-by-pixel uncertainties $\sigma(\Phi_i)$. In that case, we apply Gaussian error propagation to derive the resulting observational-only uncertainty of $Z$,
\begin{subequations}
\begin{align}
\sigma_{\rm{}obs}(Z)
    &=
    \left[
        \sum_i^{n_{\rm{}ind}}
            \left(
                \frac{\partial Z}{\partial \Phi_i} \cdot \sigma(\Phi_i)
            \right)^2
    \right]^{1/2} \\
    &=
    \frac{1}{(n_{\rm{}ind}/2)^{1/2}} \cdot
    \left[
        \sum_i^{n_{\rm{}ind}}
            \left(
                \frac{\partial \cos(2\cdot{}\Phi_i)}{\partial \Phi_i} \cdot \sigma(\Phi_i)
            \right)^2
    \right]^{1/2} \\
    &=
    \frac{1}{(n_{\rm{}ind}/2)^{1/2}} \cdot
    \left[
        \sum_i^{n_{\rm{}ind}}
            \left(
                2 \cdot \sin(2\cdot{}\Phi_i) \cdot \sigma(\Phi_i)
            \right)^2
    \right]^{1/2} \label{app-eq:Z-unc-general} \\
    &=
    \left(
        \frac{8}{n_{\rm{}ind}}
    \right)^{1/2} \cdot
    \left[
        \sum_i^{n_{\rm{}ind}}
                \sin^2(2\cdot{}\Phi_i)
    \right]^{1/2}
    \cdot \sigma(\Phi_i) \label{app-eq:Z-unc-common_noise} \\
    &=
    8^{1/2} \cdot
    \left[
        \frac{1}{n_{\rm{}ind}} \cdot
        \sum_i^{n_{\rm{}ind}}
            \sin^2(2\cdot{}\Phi_i)
    \right]^{1/2}
    \cdot \sigma(\Phi_i) \\
    &=
    8^{1/2} \cdot
    \left\langle
        \sin^2(2\cdot{}\Phi_i)
    \right\rangle
    \cdot \sigma(\Phi_i) \, .
\end{align}
\end{subequations}
Steps forward from Eq.~(\ref{app-eq:Z-unc-common_noise}) assume that the uncertainty $\sigma(\Phi_i)$ is identical for all $\Phi_i$. That idealization does not usually hold in practice, but it gives a more intuitive access to $\sigma_{\rm{}obs}(Z)$ as calculated from Eq.~(\ref{app-eq:Z-unc-general}) for the general case.

\section{Cloud-Scale Relations between Magnetic Field and Cloud Structure\label{sec-app:B-vs-density_all}}
\begin{figure}
\centerline{\includegraphics[width=0.8\linewidth]{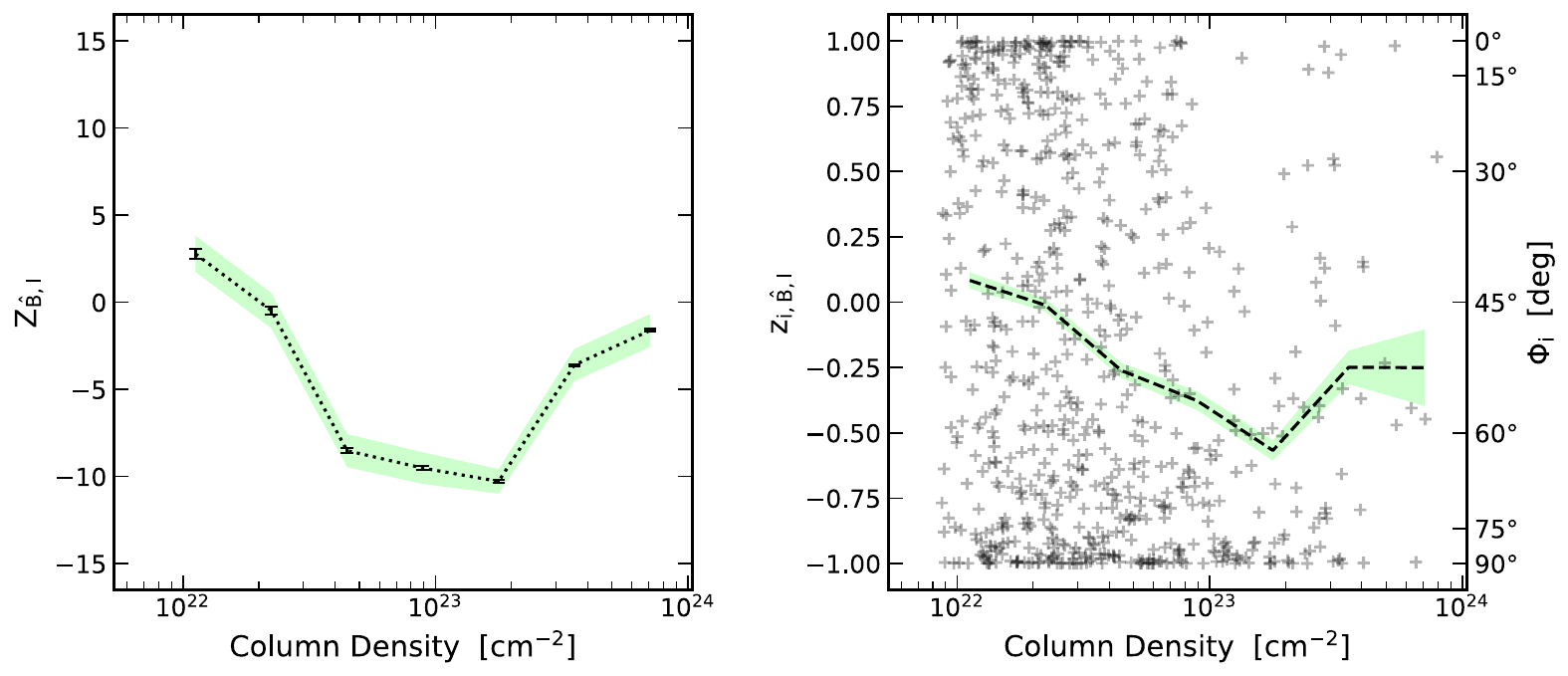}}
\caption{Relative orientation between $\hat{B}_{\rm{}pos}$ and $\hat{\nabla}_{\perp}I$ as measured for the entire cloud (Main Ridge plus all sub-filaments). Similar to Fig.~\ref{fig:relative-orientations_BGI}~(left), which we present $Z_{\hat{B},I}$ and $z_{i,\hat{B},I}$ as measured for sub-filaments and the Main Ridge, respectively.\label{fig-app:IvsB-all}}
\end{figure}

Figure~\ref{fig-app:IvsB-all} presents measurements of $Z$ and $z_i$ across the entirety of the DR21 complex, with the exception of the outflow-affected region defined in Sec.~\ref{sec:segmentation}. This should be compared to Fig.~\ref{fig:relative-orientations_BGI}~(left), which presents $Z$ and $z_i$ as measured for sub-filaments and the Main Ridge, respectively. Trends in $Z/\sigma_{\rm{}gen}(Z)$ indicate that the magnetic field is mildly aligned with elongated cloud structure at low column densities, becomes strongly perpendicular to filamentary cloud morphology at intermediate values of $N({\rm{}H_2})$, but is just mildly perpendicular to elongated substructure at the highest column densities. At low column densities, where there is signal from both sub-filaments and the Main Ridge, $Z$ falls between the trends seen for these two regions, as discussed in Sec.~\ref{sec:B-vs-density} and illustrated in Fig.~\ref{fig:relative-orientations_BGI}~(left). At high column densities, for which there is no signal from the sub-filaments, $Z$ follows the trend previously reported for the Main Ridge.

\section{Importance of Magnetic Fields for Present-Day Stability against Collapse}\label{sec-app:stability-turbulence}
Section~\ref{sec:stability-turbulence} discussed the stability of sub-filaments and the Main Ridge with respect to collapse under self-gravity. The calculations follow well-known principles and are therefore removed to this appendix.

\citet{ostriker1964:polytropes} showed that isothermal cylinders of infinite length are characterized by a critical mass-to-length ratio $(M/\ell)_{\rm{}cr,th}=2\cdot{}\sigma_{\rm{}th}^2/G$, where $\sigma_{\rm{}th}$ is the thermal sound speed and $G$ is the constant of gravity. This can be generalized to include non-thermal gas motions exhibiting a one-dimensional velocity dispersion $\sigma_{\rm{}nt}$,

\begin{equation}
\begin{split}
(M/\ell)_{\rm{}cr,kin}
    &= 2 \cdot{}
    \frac{\sigma_{\rm{}th}^2 + \sigma_{\rm{}nt}^2}{G} \\
    &= 16.43 \, M_{\odot}\,{\rm{}pc^{-1}} \cdot
    (T_{\rm{}gas} / 10 ~ {\rm{}K})
    + 464 \, M_{\odot}\,{\rm{}pc^{-1}} \cdot
    (\sigma_{\rm{}nt} / {\rm{}km\,s^{-1}})^2 \, ,
\label{eq-app:mass-per-length_cr}
\end{split}
\end{equation}
where $\sigma_{\rm{}nt}$ is calculated as
\begin{equation}
\sigma_{\rm{}nt}^2 =
    \sigma_{m,{\rm{}obs}}^2 - \sigma_{m,{\rm{}th}}^2
\label{eq-app:estimate_sigma_nt}
\end{equation}
from the observed line-of-sight gas velocity $\sigma_{m,{\rm{}obs}}$ of a molecule of mass $m$, and
\begin{equation}
\sigma_{m,{\rm{}th}}
    = (k_{\rm{}B} \cdot T_{\rm{}gas} / m)^{1/2}
    = 90.9~{\rm{}m\,s^{-1}}
        \cdot (T_{\rm{}gas}/10~{\rm{}K})^{1/2}
        \cdot (m/10\,m_{\rm{}p})^{-1/2}
\end{equation}
is the thermal velocity dispersion of that molecule (where $m_{\rm{}p}$ is the proton mass). Equation~(\ref{eq-app:mass-per-length_cr}) assumes that $\sigma_{\rm{}th}=\sigma_{m,{\rm{}th}}$ for a mean molecular weight of $m=2.33\,m_{\rm{}p}$ per free particle (Appendix~A of \citealt{kauffmann2008:mambo-spitzer}). \citet{schneider2010:dr21} observe velocity dispersions $\le{}1.3~\rm{}km\,s^{-1}$ in the DR21 Main Ridge (their Sec.~5.5.1), with values $<1~\rm{}km\,s^{-1}$ that dip to $0.5~\rm{}km\,s^{-1}$ and below prevailing throughout the outskirts occupied by the sub-filaments (their Fig.~18). They use observations of $\rm{}H^{13}CO^+$, which is characterized by $m=30\,m_{\rm{}p}$ so that Eq.~(\ref{eq-app:estimate_sigma_nt}) simplifies to $\sigma_{\rm{}nt}=\sigma_{m,{\rm{}obs}}$ for $T_{\rm{}gas}\le{}20~\rm{}K$ typically found in DR21 (\citealt{hennemann2012:dr21}, assuming that gas and dust temperatures are well coupled). For H$^{13}$CO$^+$ at $T_{\rm gas} \leq 20$\,K, Eq.~(C7) gives $\sigma_{m,\rm th} \leq 0.07$\,km\,s$^{-1}$, which is $\lesssim 15\%$ of $\sigma_{m,\rm obs}$ even in the lowest-dispersion regions of the sub-filaments and contributes $\lesssim 2\%$ to $\sigma_{\rm nt}^2$ when 
subtracted in quadrature. The thermal contribution is therefore negligible.

The observed properties give $M/\ell\gg{}(M/\ell)_{\rm{}cr,kin}$ for the Main Ridge, indicating that it is unstable with respect to collapse unless supported by strong magnetic fields. Specifically, adoption of $\sigma_{\rm{}nt}\le{}1.3~\rm{}km\,s^{-1}$ and $T_{\rm{}gas}\le{}20~\rm{}K$ in the Main Ridge yields $(M/\ell)_{\rm{}cr,kin}\le{}817\,M_{\odot}\,{\rm{}pc^{-1}}$, while $M/\ell=3,650\,M_{\odot}\,{\rm{}pc^{-1}}$ is observed (Sec.~\ref{sec:segmentation}).

The situation is less clear in the sub-filaments. Adopting $T_{\rm{}gas}\le{}20~\rm{}K$, we find $(M/\ell)_{\rm{}cr,kin}$ to range from $\le{}497\,M_{\odot}\,{\rm{}pc^{-1}}$ for $\sigma_{\rm{}nt}\le{}1~\rm{}km\,s^{-1}$ to $\le{}149\,M_{\odot}\,{\rm{}pc^{-1}}$ for $\sigma_{\rm{}nt}\le{}0.5~\rm{}km\,s^{-1}$. Given that observed values of $M/\ell$ range from $164\,M_{\odot}\,{\rm{}pc^{-1}}$ to $556\,M_{\odot}\,{\rm{}pc^{-1}}$ (Table~\ref{tab:subfilaments}), while observational constraints on $\sigma_{\rm{}nt}$ are limited, a dedicated assessment of sub-filament kinematics and masses is needed to make a conclusive statement.

Section~\ref{sec:stability-turbulence} also includes an assessment of column and volume densities in the Main Ridge. These estimates are derived using the principles documented here. We focus on regions in the Main Ridge, established by considering regions having column densities greater than the limiting column density, $N_{\rm{}H_2,lim}$. We measure the width of these regions in the East-West direction, $w$, and their mean column density, $\langle{}N_{\rm{}H_2}\rangle$. Assuming that $w$ and $\langle{}N_{\rm{}H_2}\rangle$ are similar for viewing directions in the plane of the sky, we estimate the mean density to be
\begin{equation}
\langle{}n_{\rm{}H_2}\rangle =
    32,400~{\rm{}cm^{-3}} \cdot
    \left(
        \frac{
            \langle{}N_{\rm{}H_2}\rangle
            }{
            10^{23}~\rm{}cm^{-2}
            }
    \right) \cdot
    \left(
        \frac{w}{\rm{}pc}
    \right)^{-1} \, .
\label{eq:density-from-NH2}
\end{equation}
Observed and derived properties following this analysis scheme are presented in Table~\ref{tab:stability-analysis}.

\section{Reference Accretion Rates and Velocities}\label{sec-app:accretion}
Section~\ref{sec:guided-accretion_retardation} builds on the evaluation of the  \citet{heitsch2013:infall-onto-filaments} model of accretion into an infinitely long cylinder. This appendix supports the discussion by providing a numerical evaluation of this model, as well as related calculations.

Accretion of a mass reservoir $M$ within a time $t$ requires a rate $\dot{M}_{\rm{}accr}=M/t$, which evaluates to
\begin{equation}
\dot{M}_{\rm{}accr} = 2.5\times{}10^{-3}\,M_{\odot}\,{\rm{}yr}^{-1}
    \cdot \left( \frac{M}{10^{4}\,M_{\odot}} \right)
    \cdot \left( \frac{t}{4~\rm{}Myr} \right)^{-1} \, .
\end{equation}
For the DR21 Main Ridge with a total mass of about $2\times{}10^4\,M_{\odot}$ (Sec.~\ref{sec:segmentation}) this implies rates $\dot{M}_{\rm{}accr}\sim{}5\times{}10^{-3}\,M_{\odot}\,{\rm{}yr^{-1}}$, assuming an age of 4~Myr, which is within the range of 1--5~Myr indicated by stellar colors \citep{Beerer2010:Cygnus-X}. 

A reference star formation rate for regions of dense gas is given by
\begin{equation}
\dot{M}_{\rm{}SF} = 4.6\times{}10^{-4} \, M_{\odot} \, {\rm{}yr}^{-1}
    \cdot \left( \frac{M_{\rm{}dg}}{10^{4}\,M_{\odot}} \right)
\end{equation}
\citep{lada2010:sf-efficiency}, where $M_{\rm{}dg}$ is the gas mass residing at column densities $N_{\rm{}H_2}\ge{}7\cdot{}10^{21}~\rm{}cm^{-2}$. Our mass measurements of about $2\times{}10^4\,M_{\odot}$ for the DR21 Main Ridge are obtained for regions of much higher column density, implying $M_{\rm{}dg}\gg{}2\times{}10^4\,M_{\odot}$ and thus $\dot{M}_{\rm{}SF}\gg{}10^{-3}\,M_{\odot}\,{\rm{}yr^{-1}}$. This star formation rate is 
comparable to the accretion rate 
$\dot{M}_{\rm{}accr}\sim{}5\times{}10^{-3}\,
M_{\odot}\,{\rm{}yr^{-1}}$ estimated above, 
indicating that accretion can 
plausibly sustain the observed level of star formation 
in the Main Ridge.

The free-fall velocity onto an infinitely long filament with a mass-to-length ratio $M/\ell$ is
\begin{equation}
\begin{split}
v(r) 
    &= -2 \cdot
        \left[ G \cdot \frac{M}{\ell} \cdot
            \ln(r_{\rm{}ini}/r) \right]^{1/2} \\
    &= -3.5~{\rm{}km\,s^{-1}}
    \cdot \left( \frac{M/\ell}{10^3\,M_{\odot}\,\rm{}pc^{-1}} \right)^{1/2}
    \cdot \left( \ln\left[ \frac{r_{\rm{}ini}/r}{2} \right] \right)^{1/2} \\
    &= -4.1~{\rm{}km\,s^{-1}}
    \cdot \left( \frac{M/\ell}{10^3\,M_{\odot}\,\rm{}pc^{-1}} \right)^{1/2}
    \cdot J(r/r_{\rm{}ini})
    \label{eq:free-fall-filament_velocity}
\end{split}
\end{equation}
\citep{heitsch2013:infall-onto-filaments}, assuming test particles start with zero inward velocity at the initial radius $r_{\rm{}ini}$. The function
\begin{equation}
J(r/r_{\rm{}ini}) = \left( \ln[r_{\rm{}ini}/r] \right)^{1/2}
\end{equation}
achieves values $J(x)=[1.52,1.23,1.10,0.96,0.83]$ at at $x=r/r_{\rm{}ini}=[0.1,0.2,0.3,0.4,0.5]$, making $J\approx{}1$ a good approximation for a range of situations. Given $M/\ell\approx{}4\times{}10^3\,M_{\odot}\,{\rm{}pc^{-1}}$ (Sec.~\ref{sec:segmentation}), free-fall velocities $v\approx{}-8~\rm{}km\,s^{-1}$ thus represent a meaningful characteristic infall speed for the DR21 Main Ridge. Free-fall motion following Eq.~(\ref{eq:free-fall-filament_velocity}) from a starting point $r_{\rm{}ini}$ to an end point $r$ takes a transit time $t_{\rm{}trans}=-\int_{r_{\rm{}ini}}^r{\rm{}d}r'/v(r')$, which evaluates to
\begin{equation}
\begin{split}
t_{\rm{}trans}
    &= - \frac{1}{2}
        \cdot \left( \frac{\ell}{G\cdot{}M} \right)^{1/2}
        \cdot \int_{r_{\rm{}ini}}^r
            \left[ \ln(r_{\rm{}ini}/r') \right ]^{-1/2}
            \, {\rm{}d}r' \\
    &= 4.7\times{}10^5~{\rm{}yr}
        \cdot \left( \frac{r_{\rm{}ini}}{2~\rm{}pc} \right)
        \cdot \left( \frac{M/\ell}{10^3\,M_{\odot}\,{\rm{}pc^{-1}}} \right)^{-1/2}
        \cdot H(r/r_{\rm{}ini}) \ .
\end{split}
\end{equation}
The function
\begin{equation}
H(r/r_{\rm{}ini}) =
    \int_{r/r_{\rm{}ini}}^1
            \left[ \ln(1/x) \right ]^{-1/2}
            \, {\rm{}d}x
\end{equation}
evaluates to $H\to{}\pi^{1/2}\approx{}1.77$ for $r\to{}0$, which corresponds collapse into the filament. We correspondingly define the accretion timescale to be equal to the transit time for this condition, i.e., $t_{\rm{}trans}\to{}t_{\rm{}accr}$ for $r\to{}0$.
Given 
$M/\ell\approx{}4\times{}10^3\,M_{\odot}\,
{\rm{}pc^{-1}}$ (Sec.~\ref{sec:segmentation}), 
starting points relatively nearby to the DR21 Main 
Ridge (i.e., $r_{\rm{}ini}=2~\rm{}pc$) result in 
timescales 
$t_{\rm{}accr}\approx{}4.2\times{}10^5~\rm{}yr$, 
roughly 10\% of the estimated age of the DR21 complex. 
This short timescale suggests that the sub-filaments 
observed today will merge with the Main Ridge well 
within the cloud lifetime, and that multiple 
generations of such structures could have fed the 
Ridge over the course of its evolution.

\section{Timeline for Star Formation in Sub-Filaments}
\label{sec-app:evolution-subfilaments}
Section~\ref{sec:guided-accretion_rates} discusses the accretion of sub-filaments onto the Main Ridge. Appendix~\ref{sec-app:accretion} estimates that the sub-filaments observed today will accrete into the Main Ridge on a characteristic timescale $t_{\rm{}accr}\approx{}4.2\times{}10^5~\rm{}yr$. This appendix supports Sec.~\ref{sec:guided-accretion_rates} by estimating whether sub-filaments can also collapse under self-gravity on a timescale $\lesssim{}t_{\rm{}accr}$ to form stars prior to accreting onto the DR21 Main Ridge. This calculation follows well-documented procedures and is thus removed from the main text.

The local free-fall collapse time of sub-filaments is
\begin{equation}
\begin{split}
t_{\rm ff}
    &= \left( \frac{3\cdot{}\pi}{32\cdot{}G\cdot{}\varrho} \right)^{1/2} \\
    &\approx 0.3\,{\rm{}Myr}
        \cdot \left(\frac{n_{\rm H_2}}{10^4\,{\rm cm}^{-3}}\right)^{-1/2} \, ,
\end{split}
\label{eq-app:tff}
\end{equation}
with $\varrho=\mu_{\rm{}H_2}\cdot{}n_{\rm{}H_2}$ \citep{kauffmann2008:mambo-spitzer}. This evaluates to $t_{\rm{}ff}\approx{}0.3~\rm{}Myr$ for the densities from Table~\ref{tab:subfilaments}. This timescale is similar to the timescale for sub-filaments to accrete onto the Main Ridge, $t_{\rm{}accr}\sim{}0.42~\rm{}Myr$ (Appendix~\ref{sec-app:accretion}). This suggests that sub-filaments will not undergo substantial star formation during their accretion into the DR21 Main Ridge, as this would require $t_{\rm{}ff}\ll{}t_{\rm{}accr}$, which is not the case.

This argument about the sub-filament star formation activity ignores the presence of magnetic fields. In Sec.~\ref{sec:stability-turbulence} we also argue that the Main Ridge is in a magnetically near-critical state. The same should thus also generally hold for sub-filaments, following our argument that sub-filaments have a mass-to-flux ratio comparable to the one of the Main Ridge (Sec.~\ref{sec:guided-accretion_rates}). This suggests that stars in sub-filaments would form slower than expected in the case of free-fall collapse evaluated in Eq.~(\ref{eq-app:tff}), suggesting again that sub-filaments should be mostly starless.

\end{document}